\documentclass{elsarticle}
\usepackage{fullpage}
\usepackage{graphics}
\usepackage{amssymb}
\usepackage{mathrsfs}
\usepackage{amsmath}
\usepackage{graphicx}
\usepackage{amsmath,amsfonts}
\usepackage{accents}
\usepackage{wasysym}
\usepackage{xcolor}
\usepackage{hyperref}
\usepackage{lineno}
\usepackage{setspace}
\usepackage[normalem]{ulem}

\usepackage{tikz}
\usetikzlibrary{shapes,arrows}

\newcounter{saveeqn}

\newcommand{\ppp}[2]{{\frac{\partial #1}{\partial #2}}}
\newcommand{\ppn}[3]{{\frac{\partial^{#1} #2}{\partial #3^{#1}}}}

\newcommand{\beq}{\begin{equation}}
\newcommand{\eeq}{\end{equation}}
\newcommand{\ba}{\begin{array}}
\newcommand{\ea}{\end{array}}
\newcommand{\bea}{\begin{eqnarray}}
\newcommand{\eea}{\end{eqnarray}}
\newcommand{\bbm}{\begin{bmatrix}}
\newcommand{\ebm}{\end{bmatrix}}

\newcommand{\vv}{\mathbf{v}}

\newcommand{\grad}{\bigtriangledown}

\newcommand{\scrA}{\mathcal{A}}

\newcommand{\scrE}{\mathcal{E}}

\newcommand{\rme}{\rm{e}}
\newcommand{\rmi}{\rm{i}}

\newcommand{\sn}{\mathrm{sn}}

\begin{document}

\title{On the variation of bi-periodic waves in the transverse direction} 
\author[1]{Diane M. Henderson}
\author[2]{John D. Carter}
\author[1]{Megan E. Catalano}

\cortext[cor1]{Corresponding author, dmh@math.psu.edu}
\address[1]{Department of Mathematics, Penn State University, State College, PA, 16801, USA}
\address[2]{Mathematics Department, Seattle University, Seattle, WA, 98122, USA}

\begin{abstract}

\bigskip

    \noindent Weakly nonlinear, bi-periodic patterns of waves that propagate in the $x$--direction with amplitude variation in the $y$--direction are generated in a laboratory. The amplitude variation in the $y$--direction is studied within the framework of the vector (vNLSE) and scalar (sNLSE) nonlinear Schr\"odinger equations using the uniform-amplitude, Stokes-like solution of the vNLSE and the Jacobi elliptic sine function solution of the sNLSE.  The wavetrains are generated using the Stokes-like solution of vNLSE; however, a comparison of both predictions shows that while they both do a reasonably good job of predicting the observed amplitude variation in $y$, the comparison with the elliptic function solution of the sNLSE has significantly less error
when the ratio of $y$--wavenumber to the two-dimensional wavenumber is less than about 0.25. For ratios between about 0.25 and 0.30 (the limit of the experiments) the two models have comparable errors.
When the ratio is less than about 0.17, agreement with the vNLSE solution requires a third-harmonic term in the $y$--direction, obtained from a Stokes-type expansion of interacting, symmetric wavetrains. There is no evidence of instability growth in the $x$--direction, consistent with the work of Segur and colleagues, who showed that dissipation stabilizes the modulational instability. Finally, there is some extra amplitude variation in $y$, which is examined via a qualitative stability calculation that allows symmetry breaking in that direction.
    \begin{center}
        {\it{We dedicate this paper to our friend and colleague, Harvey Segur.}}
    \end{center}
\end{abstract}

\maketitle

\section{Introduction}
\label{intro}
We consider the evolution of weakly nonlinear, bi-periodic surface water waves on deep water generated in a laboratory wavetank.  A directional wavemaker creates the wave patterns, which propagate in the $x$--direction and have crests with amplitudes that
vary in the $y$--direction.
The wavefields are generated by programming the wavemaker to produce obliquely interacting waves with a prescribed $y$-wavelength.
These waves may be modeled (for example) by the vector nonlinear Schr\"odinger equation (vNLSE), 
e.g. \cite{roskes76}, which is a system of two coupled nonlinear partial differential equations (PDEs) that describe the evolution of the amplitude envelopes of the two obliquely interacting wavetrains;
or the scalar nonlinear Schr\"odinger equation (sNLSE),
e.g. \cite{as}, which describes the evolution of the amplitude envelope of a single wavetrain that may have $x$ and $y$ wavenumbers. 
The sNLSE assumes that the single wavetrain is weakly two-dimensional (in the horizontal). 
Its derivation requires that $\epsilon_y$, the ratio of the $y$--wavenumber to the two-dimensional wavenumber, is small; that is,
$\epsilon_y \ll 1$. 
The vNLSE allows for two interacting wavetrains and makes no such constraint on $\epsilon_y$. 
These two models have exact solutions that are relevant for our experiments.
The vNLSE has a traveling-wave, bi-periodic, Stokes-type solution for symmetric waves that corresponds to a pattern that propagates with uniform amplitude in the $x$--direction and has a cosine-type amplitude variation along the crests in the $y$--direction.
The sNLSE has a traveling-wave, Jacobi elliptic sine (sn) function solution that
corresponds to a pattern that propagates with uniform amplitude in the $x$--direction and has a sn-type amplitude variation along the crests in the $y$--direction.
We compare measurements of the $y$--variation of amplitudes to the predictions from these two solutions. 

The waves in the experiments are generated using the Stokes-type solution of the vNLSE. 
Nevertheless, we find that for $\epsilon_y \lessapprox 0.25$, the sNLSE is the better model; the error in comparisons of measured amplitude variation in the $y$-direction is smaller for the sn solution of the sNLSE than is the Stokes-type solution of the vNLSE.
The agreement with the vNLSE solution is improved by adding a forced, higher-order (in the small parameter, $\epsilon$, that measures weak nonlinearity) term to the Stokes-type solution, but the error is still larger than that of the sn solution of the sNLSE.
For $0.25 \lessapprox \epsilon_y \lessapprox 0.30$, the errors between measurements of the amplitude variation in the $y$--direction and the solutions of sNLSE and vNLSE are comparable whether or not the higher-order (in nonlinearity) term is added to the Stokes-type solution
of the vNLSE. ($\epsilon_y \approx 0.30$ is the limit for our experiments.)
Predictions from both equations are for amplitude variations that are periodic in $y$; 
however, measurements show some extra amplitude variation in $y$. 
One explanation for the extra variation might be the stability to symmetry-breaking perturbations in the $y$--direction that are not required to grow in time (or correspondingly in $x$). We consider this possibility with a qualitative stability calculation of the sn-function solution of the sNLSE.

Many authors have considered the stability of patterns described by the Stokes-type solution of vNLSE, including 
\cite{bn67} and \cite{roskes76}, who used vNLS-type equations;
\cite{dd91}, who allowed for higher-order terms like those in the Dysthe \cite{dysthe79} equation; and
\cite{ik94}, \cite{bski95}, and \cite{leblanc09}, who allowed for two-dimensional perturbations.
Stability of these solutions within the vNLSE framework has also been investigated by, for example,
\cite{oos06} and \cite{skms06}. A numerical investigation of stability of this type of pattern using coupled
deep-water Boussinesq equations was conducted by \cite{fmb99}.
The sNLSE has many traveling-wave solutions that can be expressed in terms of Jacobi elliptic functions (e.g. \cite{cc00} and \cite{cc00b}).
Deconinck \& Lovit \cite{dl10} showed that these solutions provide orthonormal bases for square integrable functions with periodic
boundary conditions and proposed using such a nonlinear basis rather than, for example, a Fourier basis, because the nonlinear
basis requires fewer modes than does the Fourier basis. 
Consistent with this idea is the result herein that
a single sn solution of sNLSE agrees better with experiments than does the Fourier solution resulting from vNLSE
corrected by the addition of the third-harmonic term.
Importantly, the nonlinear basis proposed in \cite{dl10} has phase information built in that is
lost when using a Fourier basis with a random phase approximation. 
However, as is true for the Stokes-type solution of the vNLSE, the stability of these elliptic function solutions has been investigated by \cite{cs03} and \cite{cd06},
who showed that 
every one-dimensional (trivial-phase) traveling-wave solution to the sNLSE is unstable with respect to two-dimensional perturbations.
 
 All of these works, which consider inviscid dynamics, either in a vNLSE or sNLSE context,
 show that waves with two-dimensional surface patterns are unstable in deep-water. However, Segur and colleagues
 showed that the inclusion of dissipation of a particular form (Rayleigh-type) changes this result. 
Segur {\emph{et al.}} \cite{bf05}
showed that including dissipation in the one-dimensional sNLSE equation stabilizes the modulational instability
(the Benjamin-Feir instability, \cite{bf67}); that is, dissipation can stop the growth of a perturbation before nonlinear effects
become large enough to play a role. 
Henderson and Segur \cite{hs13}, showed that stabilization by dissipation may have applications to the stabilization of ocean swell, which
have been observed to propagate stably across the Pacific Ocean, \cite{wap66}. 
The stabilizing effect is not restricted to waves propagating in one-dimension. 
Segur {\emph{et al.}} \cite{hsc10}  
further showed that dissipation stabilizes bi-periodic wave patterns, such as those being studied herein. 
Consistent with the results of Segur and colleagues, our wave patterns 
do not show evidence of modulational instability in the $x$--direction. 
Also consistent with their results is the observation that the extra amplitude variation in the $y$--direction (discussed above) decreases in $x$, the direction of propagation. 

An outline of the remainder of the paper is as follows.  
In \S \ref{theory} we outline the derivation of the vNLSE and sNLSE, and present the Stokes-type solution of the vNLSE and the sn solution of the sNLSE.  
Following \cite{cd06} we consider a qualitative stability calculation of the sn solution with respect to perturbations in the $y$--direction.
The experimental apparatus and procedures are described in \S{\ref{experiments}}. Results are presented in \S{\ref{results}} and are summarized in \S{\ref{summary}}.

\section{Theoretical Considerations}
\label{theory}

Two obliquely interacting waves in deep water that are weakly nonlinear can be modeled by the vNLSE (vector nonlinear Schr\"odinger equations), which are two coupled PDEs for the envelopes of each of the interacting waves and by the sNLSE (scalar nonlinear Schr\"odinger equation), which is a single PDE that models a wavetrain propagating in the $x$--direction with amplitude variation in the $y$--direction. 
We begin by presenting an outline of the derivation of these equations and the solutions relevant to our experiments. 

The vNLSE and sNLSE for the evolution of water wave envelopes are derived from the Stokes boundary value problem \cite{stokes} for waves on an inviscid fluid with irrotational motions. Here we consider a domain of infinite horizontal extent and depth, and allow for the restoring forces of gravitation and capillarity. Then the irrotational velocity field, $\vv(x, y, z, t) = \grad \phi(x, y, z, t)$ and the free surface displacement $\eta(x, y, t)$ are determined by the statement of conservation of mass with boundary conditions, 
\begin{subequations}
\bea
\Delta \phi = 0 ~~~~~~ {\rm in} ~~~~ -\infty < z < \eta(x,y,t),&&\!\!\!\!\!\!\!\! 
-\infty < x,y < \infty\\
\ppp{\eta}{t}
+ \nabla \phi \cdot \nabla \eta 
 -\ppp{\phi}{z}=0 & {\rm on} & z=\eta(x,y,t)\\
\ppp{\phi}{t} + \frac{1}{2} |\nabla \phi|^2  + g~\eta  = \sigma \frac{(1+\frac{\partial\eta}{\partial y})\frac{\partial^2\eta}{\partial x^2}+(1+\frac{\partial\eta}{\partial x})\frac{\partial^2\eta}{\partial y^2}-2\frac{\partial\eta}{\partial x}\frac{\partial\eta}{\partial y}\frac{\partial^2\eta}{\partial x\partial y}}{(1+(\frac{\partial\eta}{\partial x})^2+(\frac{\partial\eta}{\partial y})^2)^{3/2}}& {\rm on} & 
z = \eta(x,y,t)\\
\grad \phi=0 & {\rm on} & z\rightarrow -\infty,
\eea
\label{thebvp}
\end{subequations}
\noindent where $g$ is the acceleration due to gravity and $\sigma$ is the coefficient of kinematic surface tension.
Following the procedure of the method of multiple scales (e.g. see \cite{as}, pp. 251-252 for a general procedure and pp. 317-323 for an the application to deep-water waves), we expand the free surface displacement and velocity potential in a small parameter, $\epsilon$,
such that
\begin{subequations}
\beq
\eta(x,y,t) = \sum_{j=1}^\infty \epsilon^j \eta_j(x,y,t, X, Y, T, T_2),  
\label{etaexpand}
\eeq
\beq
\phi(x,y,z,t) = \sum_{j=1}^\infty \epsilon^j \phi_j(x,y,z,t, X, Y, Z, T, T_2) ,
\eeq
\label{expandall}
\end{subequations}
\noindent where the variables are functions of time, $t$, spatial variables, $\{x, y, z\}$, as well as the slow space and time scales,
\beq
X = \epsilon x, ~~~Y = \epsilon y, ~~~Z = \epsilon z, ~~~T = \epsilon t, ~~~{\rm and} ~~~ T_2 = \epsilon^2 t.
\label{scales}\eeq
Use ({\ref{expandall}}) in ({\ref{thebvp}}) 
to obtain
an ordered sequence of inhomogeneous, linear boundary--value problems for the $\eta_j$ and
$\phi_j$. We choose the appropriate first-order solution in \S{\ref{theory-vnlse}} and \S{\ref{theory-snlse}} to derive either the vNLSE
or the sNLSE.
\subsection{vNLSE}
\label{theory-vnlse}
For the case of two wavetrains propagating at an oblique angle with complex amplitudes, $A/B$;
$x$-wavenumbers, $k_{A/B}$;
$y$-wavenumbers, $l_{A/B}$; and frequencies, 
$\omega_{A/B}$;
we follow \cite{hhs05} and express the first-order in $\epsilon$ term for the surface displacement ({\ref{etaexpand}}) 
as
\beq
\eta_1(x, y, t) =  \frac{{\rmi}}{2} \Big(A(X, Y, T, T_2) {\rme}^{{\rmi}\theta_A} - A^* {\rme}^{-{\rmi}\theta_A}\Big) +  
 \frac{{\rmi}}{2} \Big(B(X, Y, T, T_2)  {\rme}^{{\rmi}\theta_B} - B^* {\rme}^{-{\rmi}\theta_B}\Big),
\label{wavesAB}
\eeq
where $\theta_{A/B} = k_{A/B} x + l_{A/B} y - \omega_{A/B} t$, the amplitudes depend on the slow variables, and the asterisk represents the complex conjugate. 
Using ({\ref{wavesAB}}) the $\mathcal{O}(\epsilon)$ system results in the linear dispersion relation between
the $x$ and $y$ wavenumbers and the frequencies,
\beq
\omega_{A/B}^2=g \kappa_{A/B} + \sigma \kappa_{A/B}^3,
\label{disper}
\eeq
where
$\kappa_{A/B} = \sqrt{k_{A/B}^2 + l_{A/B}^2}$.
At $\mathcal{O}(\epsilon^2)$ the amplitudes satisfy linear transport equations,
\begin{subequations}
\beq
\ppp{A}{T} + \mathbf{C}_g^A \cdot \grad_\perp A =0,
\eeq
\beq
\ppp{B}{T} + \mathbf{C}_g^B \cdot \grad_\perp B=0,
\eeq
\label{transport}
\end{subequations}
where $\mathbf{C}_g^{A/B} = (c_{A/B}, d_{A/B})$ are the group velocity vectors of the $A$ and $B$ waves and $\grad_\perp = (\ppp{}{X}, \ppp{}{Y})$. The group velocities components are given by 
\begin{subequations}
\beq
c_{A/B} =\frac{\partial \omega_{A/B}}{\partial {k_{A/B}}}=\frac{k_{A/B}(g+3\sigma\kappa^2_{A/B})}{2\kappa_{A/B}\omega_{A/B}},
\eeq
\beq
d_{A/B} = \frac{\partial \omega_{A/B}}{\partial {l_{A/B}}}=\frac{l_{A/B}(g+3\sigma\kappa^2_{A/B})}{2\kappa_{A/B}\omega_{A/B}},
\eeq
\label{group}
\end{subequations}
where $\omega_{A/B}$ is defined in equation ({\ref{disper}}). For the symmetric case, such as those considered herein for which
$k_A = k_B =: k$,  $l_A = - l_B =: l$, $ \omega_A = \omega_B =: \omega$, and $ \kappa_A = \kappa_B =: \kappa$,
the group velocities are related by
$c_A = c_B$ and
$d_A= - d_B$.

The $\mathcal{O}(\epsilon^3)$ system results in evolution equations that describe the slow modulations of the carrier-wave amplitudes. These equations are the
vector nonlinear Schr\"odinger equations (vNLSE),
which are given by
\bea
{\rmi} \Big( \ppp{A}{T} + c_A \ppp{A}{X} + d_A \ppp{A}{Y} \Big) + \epsilon \Big[\lambda_A \ppn{2}{A}{X}  + \mu_A \ppn{2}{A}{Y} 
                            + \gamma_A \frac{\partial^2 A}{\partial X \partial Y} 
               + \chi_A |A|^2 A + \zeta_{AB} |B|^2 A \Big] &=& 0, \nonumber \\
{\rmi} \Big( \ppp{B}{T} + c_B \ppp{B}{X} + d_B \ppp{B}{Y} \Big) + \epsilon \Big[\lambda_B \ppn{2}{B}{X}  + \mu_B \ppn{2}{B}{Y} 
                            + \gamma_B \frac{\partial^2 B}{\partial X \partial Y} 
               + \chi_B |B|^2 B + \zeta_{BA} |A|^2 B \Big] &=& 0.
\label{vnlse}
\eea
The coefficients are given by the group velocities ({\ref{group}}) and by second-order derivatives using
({\ref{disper}}), so that
\begin{subequations}
\beq
\lambda_{A/B} = \frac{1}{2} \frac{\partial^2 \omega_{A/B}}{\partial {k_{A/B}}^2}=\frac{-g^2(k_{A/B}^2-2l_{A/B}^2)+2g\sigma\kappa_{A/B}^2(3k_{A/B}^2+4l_{A/B}^2)+3\sigma^2\kappa_{A/B}^4(k_{A/B}^2+2l_{A/B}^2)}{8\kappa_{A/B}^3(g+\sigma\kappa_{A/B}^2)\omega_{A/B}},
\eeq
\beq
\mu_{A/B} = \frac{1}{2}\frac{\partial^2 \omega_{A/B}}{\partial {l_{A/B}}^2}=-\frac{k_{A/B}l_{A/B}(3g^2+2g\sigma\kappa_{A/B}^2+3\sigma^2\kappa_{A/B}^4)}{4\kappa_{A/B}^3(g+\sigma\kappa_{A/B}^2)\omega_{A/B}},
\eeq
\beq
\gamma_{A/B} = \frac{\partial^2 \omega_{A/B}}{\partial k_{A/B}\partial l_{A/B}}=\frac{-g^2(-2k_{A/B}^2+l_{A/B}^2)+2g\sigma\kappa_{A/B}^2(4k_{A/B}^2+3l_{A/B}^2)+3\sigma^2\kappa_{A/B}^4(2k_{A/B}^2+l_{A/B}^2)}{8\kappa_{A/B}^3(g+\sigma\kappa_{A/B}^2)\omega_{A/B}}.
\eeq
\end{subequations}
For the symmetric case,  $\lambda_A = \lambda_B=:\lambda$, $\mu_A = \mu_B=:\mu$, and 
$\gamma_A =- \gamma_B=:\gamma.$
The coefficients of the nonlinear terms are coupling coefficients; $\chi_{A/B}$ are self-coupling coefficients, and $\zeta_{AB/BA}$ are cross-coupling coefficients. 
For the symmetric case, $\chi_A = \chi_B =:\chi$, and $\zeta_{AB} = \zeta_{BA} =: \zeta$, where
\begin{subequations}
  \beq
  \chi=-\frac{\kappa^2(8g^2+g\sigma\kappa^2+2\sigma^2\kappa^4)\omega}{16(g-2\sigma\kappa^2)(g+\sigma\kappa^2)},
  \label{chi-vnlse}
  \eeq
  \beq
  \zeta  = - \frac{\omega(\kappa)}{\kappa^2} \frac{(k^5 - k^3 l^2 - 3 k l^4 -2 k^4 \kappa + 2 k^2 l^2 \kappa + 2 l^4 \kappa)}{k - 2 \kappa}.
  \label{zeta-vnlse}
  \eeq
\end{subequations}
The formula for $\zeta$ given in equation (\ref{zeta-vnlse}) assumes that there is no surface tension, i.e.~$\sigma=0$.  The formula for $\zeta$ for nonzero surface tension is presented in \ref{append} due to its length.  Versions of these VNLSE coefficients without surface tension were originally presented in~\cite{hhs05,oos06}, though both of those works included minor typos.  The uniform-amplitude solution to the vNLSE ({\ref{vnlse}}) is obtained by setting the $X$ and $Y$ derivatives to zero and solving the resulting coupled first-order in time ODEs (see also \cite{hhs05}). For the symmetric case, the solution is
\beq
A(T) = B(T) = A_0  {\rme}^{{\rmi} \epsilon (\chi + \zeta) |A_0|^2  T},
\label{vsoln}
\eeq
where $A_0$ is a complex constant. Setting $a_0 = 2\epsilon |A_0|$, the first-order surface displacement may be written as
\bea
 \eta_1(x, y, t) &=&   \frac{a_0}{2} \cos(k  x + l y - \omega t) +  \frac{a_0}{2} \cos(k x - l y - \omega t),
    \nonumber \\
               &=& a_0 \cos(l y) \cos(k x - \omega t).
\label{wavesAA}
\eea
Equation ({\ref{wavesAA}}) shows that the interacting waves for the symmetric case create a pattern that propagates in the $x$-direction with an amplitude that varies sinusoidally in the $y$-direction. 
We note that the amplitude and shape of the wave pattern does not involve the coefficients in the vNLSE. Instead, these coefficients appear in
the first-order in $\epsilon$ correction to the wave phase. For experiments, we compare predictions and measurements of the shape of the wave patterns not the wave phases. Thus the values of the coefficients are not needed in \S{\ref{resultsvnlse}} where we compare predictions from vNLSE with experiments.

In deriving the vNLSE, one also derives the higher-order corrections to the carrier wave pattern, that is,
expressions for $\eta_2$ and $\eta_3$. See, e.g.  \cite{hhs05} for details. Because they are higher-order in $\epsilon$ than is $\eta_1$, they are  much smaller in amplitude than $\eta_1$, and because they are terms tied to the amplitude of the $\eta_1$, that is, they do not evolve or grow independent of $\eta_1$, they
are typically not important in describing the observed pattern of obliquely interacting waves. 
Nevertheless,
in trying to generate experimentally the permanent form wave pattern described by ({\ref{wavesAA}}), Hammack {\it{et al.}} \cite{hhs05} found three features about the pattern that were unsteady.  
Furman \& Madsen \cite{fm06} showed that the cause of the unsteadiness was due to the neglect of one higher-order term in the wavemaker forcing. They used ({\ref{wavesAA}}) to initiate numerical simulations of wave pattern evolution as had the investigators in \cite{hhs05} to program their physical wavemakers and found the same unsteadiness. When Furman \& Madsen changed the  initialization to include a specific third-order term, the unsteadiness disappeared. Henderson {\it{et al.}} \cite{hps06} showed that if they included this same third-order term in their wavemaker forcing, then the unsteadiness in their experiments disappeared. To see how this one term becomes important, consider the third-order solution,
\bea
\eta_3(x, y, t) &=& \frac{a_0^3 k^2}{2} [b_{11}\cos(l y) \cos(k x - \omega t) + 
b_{13}\cos(3 l y) \cos(k x - \omega t) \nonumber \\
  &+& b_{31}\cos(l y) \cos(3 k x - 3 \omega t) +
b_{33}\cos(3 l y) \cos(3 k x -  3 \omega t)],
\label{third}
\eea
where the $b_{nm}$ are coefficients that depend on the wavenumbers and frequencies of the carrier wave.
The $b_{11}$ term is not important; it has the same spatial and temporal dependence as the carrier wave and so provides 
an $\mathcal{O}(\epsilon^3)$ correction to the wave pattern amplitude. For the other terms, note that
none of $( k, 3 l, \omega)$, $(3 k,   l, 3 \omega)$ or $(3k, 3 l, 3 \omega)$ satisfy the free-wave dispersion relation given in equation ({\ref{disper}}). 
These are all forced waves that are part of the solution to the fully nonlinear boundary-value problem, ({\ref{thebvp}}). 
They are bound to the first-order solution.
But, Furman \& Madsen \cite{fm06} showed that neglecting these bound waves in the wave generation procedure leads to 
spurious free waves of the form,
\bea
\eta_3^{free}(x, y, t) &=&- \frac{a_0^3 k^2}{2} [ b_{13}\cos(3 l y) \cos(k_{13} x - \omega t) \nonumber \\
  &+& b_{31}\cos(l y) \cos(3 k_{31} x - 3 \omega t) +
b_{33}\cos(3 l y) \cos(3 k_{33} x -  3 \omega t)],
\label{thirdfree}
\eea
where, importantly, all of $( k_{13}, 3 l, \omega)$, $(k_{31},   l, 3 \omega)$ and $(k_{33}, 3 l, 3 \omega)$ do
satisfy the free wave dispersion relation. 
The $b_{31}$ and $b_{33}$ terms are unimportant; they are third harmonic in time and close to being third harmonic in the $x$-direction; they have a small effect on the overall amplitude and an asymmetry in the sinusoidal shape of the individual waves.  
However, the $b_{13}$ term can be important. It has the same frequency as the carrier waves 
but with an $x$-wavenumber that is not equal to $k$. This mismatch in $x$-wavenumbers, means that the spurious free wave travels at a slightly different speed than the carrier wave instead of being bound to it. The difference between the $k$ and the $k_{13}$ causes an
unsteady modulation, a beat, on the wave pattern in the $x$-direction, an unsteady curving of the crestlines, and unsteady dips and peaks in the $y$-shape of the wave pattern. Furman \& Madsen \cite{fm06} showed that including the $b_{13}$ term in the forcing in their numerics removed all three unsteady features. Henderson {\it{et al.}} \cite{hps06} showed that including that term in their wavemaker forcing removed all three unsteady features in their experiments. Therefore, we include the term,
\beq
\frac{a_0^3 k^2}{2} b_{13} \cos(3 l y) \cos(\omega t- k x),
\label{3rd}
\eeq
 in our wavemaker forcing (see \S{\ref{experiments}}) to produce as steady as possible wave patterns of interacting, symmetric wavetrains. In addition, since we are concerned with the shape of the waves in the $y$-direction, we also include this third-harmonic-in-$y$ term ({\ref{3rd}}) in our expression for $\eta$ when comparing with our measurements of the $y$-amplitude variations (see \S{\ref{resultsvnlse}}). 
The coefficient is given in \cite{hts79} and \cite{fm06}, and is
\bea
b_{13}&=& (9 \omega_0^4 - 6 + 2 \omega_0^4)/16 - S^2(3 \omega_0^{-8} +5) /16 + C^2(3 \omega_0^{-8} + 1)/16 \nonumber \\
    &+& [16 (\nu \tanh(\nu k h ) - \omega_0^2)]^{-1} [(-3 \omega_0^{-6} + 8 \omega_0^{-2} - 3 \omega_0^2 + 2 \omega_0^6)\nonumber \\
    &+& S^2 (-6 \omega_0^{-6} + 4 \omega_0^{-2} - 10 \omega_0^2) + C^2(6 \omega_0^{-6} - 4 \omega_0^{-2} - 2 \omega_0^2) \nonumber \\
    &+& 4 C^2(S^2 - C^2)\omega_0^{-2}], 
\label{eqnb13} 
\eea
with 
$\omega_0 = \sqrt{(1 + \sigma \kappa^3/g \kappa) \tanh(\kappa h)}$ ($\sigma=0$ in \cite{hts79} and \cite{fm06}); 
$h$ is the water depth, but for our experiments, $\tanh(\kappa h)\rightarrow 1$;
$S=\sin\hat\theta$; $C=\cos\hat\theta$;
$\hat\theta=\tan^{-1}(k/l)$; and $\nu = (S^2 + 9 C^2)^{1/2}$. 

In summary, from the perspective of the vNLSE model, the surface displacement may be modeled by
\beq
\eta(x, y, t)  \approx a_0 \cos(l_n y) \cos(k x - \omega t) + a_3 \cos(3 l_n y) \cos(k x -\omega t ),
\label{sdvnlse}
\eeq
where 
\beq
a_3 = \frac{a_0^3 k^2}{2} b_{13}.
\label{a3}
\eeq
For waves in a wavetank of width $W$, so that $0 \le y \le W$, in order to satisfy the boundary conditions of no-flow through the vertical side-walls, $\partial \phi/\partial y = 0$ on $y = 0, W$, the $y$-wavenumber is digitized to be $l= l_n = n \pi / W$, $n = 1, 2, \dots$, .
The mode number, $n$, corresponds to the number of nodal lines in the $y$--direction, that is, the number of lines parallel to the
$x$-axis for which there is zero surface displacement.


\subsection{sNLSE}
\label{theory-snlse}

Although in the experiments (\S{\ref{experiments}}) we program the wavemaker with ({\ref{sdvnlse}}), we show in \S{\ref{results}} that the resulting wave patterns are better described by a different model of two-dimensional patterns of waves. In this section, we present that model.

The surface displacement, $\eta(x, y, t)$, due to a single, traveling wavetrain whose amplitude varies slowly in both the  $x$- and $y$-directions  
may be described at leading
order in $\epsilon$ to be
\beq
\eta_1(x, y, t) =  {\rmi}\Big(\scrA(\xi, Y, T_2) {\rme}^{{\rmi}\theta} - \scrA^*{\rme}^{-{\rmi}\theta}\Big),
\label{onewave}
\eeq
where the amplitude may depend on the slow scales given in (\ref{scales}) using the (slow) translating variable,
$\xi=  (X - C_g T)$ in which $C_g$ is the group velocity. The phase is $\theta = k x + l y - \omega t$, where
$\{k, l, \omega \}$ are the $x$-wavenumber, $y$-wavenumber and frequency of the carrier wave.
A measure of two-dimensionality of the wave pattern is given by $\epsilon_y = l/\kappa$, where $\kappa = \sqrt{k^2 + l^2}$. 
The derivation of the vNLSE makes
no approximation on the size of $\epsilon_y$, but the derivation of sNLSE requires that $\epsilon_y \ll 1$. 

Following the procedure discussed in \S{\ref{theory-vnlse}} one finds the dispersion relation at  $\mathcal{O}(\epsilon)$,
\beq
\omega(\kappa) = \sqrt{ g \kappa + \sigma \kappa^3}.
\label{dispersingle}
\eeq
At $\mathcal{O}(\epsilon^2)$, one obtains the linear transport equation similar to ({\ref{transport}}), which suggests the use of $\xi$.
At $\mathcal{O}(\epsilon^3)$
the evolution of the amplitude envelope of the carrier wave may be described by the the scalar nonlinear Schr\"odinger equation (sNLSE),
\beq
{\rmi} A_{T_2} + \lambda A_{\xi\xi} + \mu A_{Y Y } + \chi |A|^2|A =0,
\label{snlse}
\eeq
where expressions for the coefficients, $\lambda,~\mu$, and $\chi$ are given in \cite{as}. 
Again, $\lambda= \frac{1}{2} \frac{\partial^2 \omega}{\partial {k}^2}$, and 
$\mu = \frac{1}{2}\frac{\partial^2 \omega}{\partial {l}^2}.$
Dimensionalizing the results given in \cite{as} (p. 320), correcting a typo, and using the approximation that
$l/\kappa \ll 1$,
the self-coupling coefficient, including surface tension can be written as
\beq
\chi =- \frac{(g \kappa)^{3/2}} {4 \omega^2} \kappa^2 \Bigg( 2 - \frac{3 \tilde{T}}{1 + \tilde{T}}\Bigg),
\label{chisnlse}
\eeq
where $\tilde{T} = k^2\sigma/g$. 
The typo (personal communication from Harvey Segur), is the $(2-\sigma)$ factor in the first line of (4.3.26) in
\cite{as} (p. 320). The 2 should be a 3.

The sNLSE admits a large class of one-dimensional traveling wave solutions with trivial phase. Carter \& Segur \cite{cs03} 
list them and examine their stability. Here we are interested in the solution that may model obliquely interacting waves, so we consider
the Jacobi elliptic sine (sn) function that provides an amplitude variation in the $Y$--direction,
\beq
A = A_{sn} \sn\Big[b Y , m\Big] {\rme}^{ {\rmi} r T_2},
\label{exact}
\eeq
where $0 < m < 1$ is the elliptic modulus, $b$ is the $Y$-wavenumber, and
\beq
A_{sn} = \sqrt {\frac{2 \mu m}{\chi} }b, ~~~~~ r = - \mu (1 + m) b^2.
\label{snA0r}
\eeq 
Then the leading-order surface displacement may be modeled by
\beq
\eta(x, y, t) = a_{sn} ~\sn\Big[c \Big(y + \frac{W}{2n}\Big), m\Big] \cos(k x - \omega t),
\label{exactsurface}
\eeq
where
\beq 
a_{sn}= 2 \epsilon A_{sn}, \hspace{1cm} c = \epsilon b  = 4 K(m)/L_y,
\label{a0c}
\eeq
are the amplitude and $y$-wavenumber. The $K(m)$ is the complete elliptic integral of the first kind, 
and $L_y = 2 W/n$ is the $y$-wavelength, where $n$ is the same mode number as used in \S{\ref{theory-vnlse}}. 
The shift in $y$ is to ensure that the envelopes
have antinodes at the sidewalls of the tank.
Carter \& Segur \cite{cs03} and Carter \& Deconinck \cite{cd06} considered the stability of elliptic function solutions
of ({\ref{snlse}}) to perturbations with wavenumbers in the direction orthogonal to that of the elliptic function's dependence. 
Our experiments show that the carrier wave amplitudes are fairly uniform in the $x$--direction as expected,
since it is known (see for example, \cite{bf05} and \cite{hsc10}) that dissipation stabilizes modulational instabilities in the $x$--direction.
However, the experiments show extra amplitude variation in the $y$--direction. As a possible explanation for the extra variation, we consider the stability of the sn solution to perturbations that have a $Y$-dependence. For the sake of generality, we allow the perturbation to have a periodicity in 
$\xi$ and a growth rate in $Y$. 
To this end, we consider a perturbed solution to ({\ref{snlse}}) of the form
\beq
A(\xi, Y, T_2) = \Big( A_{sn} {\sn} \Big[b Y , m\Big]     + \hat\epsilon u(\xi, Y, T_2) + {\rmi} \hat\epsilon v(\xi, Y, T_2) \Big) {\rme}^{ {\rmi} r T_2},
\label{snApert}
\eeq
where $u(\xi, Y, T_2)$ and $v(\xi, Y, T_2)$ are real-valued functions and $0<\hat\epsilon\ll 1$ is a small parameter.  Substituting ({\ref{snApert}}) into ({\ref{snlse}}) and linearizing gives PDEs for $u$ and $v$ that have constant coefficients in $\xi$ and $T_2$, but not in $Y$. Without loss of generality, assume
\bea
 u(\xi, Y, T_2)&=&U(Y) {\rme}^{ {\rmi} \rho \xi + \Omega T_2}+c.c.,\nonumber \\
v(\xi, Y, T_2)&=&V(Y) {\rme}^{ {\rmi} \rho \xi + \Omega T_2}+c.c.,
\label{snfpertV}
 \eea
where $U(Y)$ and $V(Y)$ are real-valued functions, $\rho$ is a real number that represents the perturbation wave number in the $\xi$--direction, $\Omega$ is a complex number that determines the behavior of the perturbations in $T_2$, and $c.c.$ represents complex conjugate.  The resulting ordinary differential equations for $U$ and $V$ are
\bea
 U^{\prime\prime}(Y) + \Big[ b^2  (1  + m ) - \frac{ \lambda \rho^2}{\mu}  - 6 b^2 m ~ {\sn}^2(b Y, m)  \Big] U(Y) 
           - \frac{\Omega}{\mu} V(Y) &=& 0 \nonumber \\
V^{\prime\prime}(Y) + \Big[ b^2(1 + m) - \frac{\lambda \rho^2 }{\mu}  - 2 b^2 m ~{\sn}^2(b Y, m) \Big] V(Y)  
          + \frac{\Omega}{\mu} U(Y)&=& 0. 
\label{eqnUV}
\eea
We do not do an exhaustive stability analysis to allow for spectra in 
$\rho$ and $\Omega$.
Instead, we compute the shape of the perturbation in the (bounded) $Y$--direction, keeping in mind the following.
In the experiments, we do not observe modulations in the $\xi$--direction, 
so we set $\rho=0$ in  ({\ref{eqnUV}}). Similarly, we do not see perturbations in the $y$--direction grow in the direction of propagation, so we set
$\Omega=0$ in ({\ref{eqnUV}}).
Figure~{\ref{figsnstab} shows a numerical solution of ({\ref{eqnUV}}) on the (bounded) $Y$ domain, for the perturbation amplitude
and provides a qualitative explanation for the observed left-right asymmetry. 
The calculation shown here 
is for homogeneous, Neumann boundary conditions, $\{U, V\}^\prime(Y) = 0$ at $Y = \{0, W\}$, and corresponds to a $4$-Hz carrier wave
so that 
$\lambda = -5.965,~\mu = 16.666,~\epsilon b = 0.173,~\chi =4.519$. The mode number is $n=4$, and $m=0.997$
to correspond to the $n=4$ experiment discussed in \S{\ref{results}}.
\begin{figure}[ht]
\center{\includegraphics[width=3in]{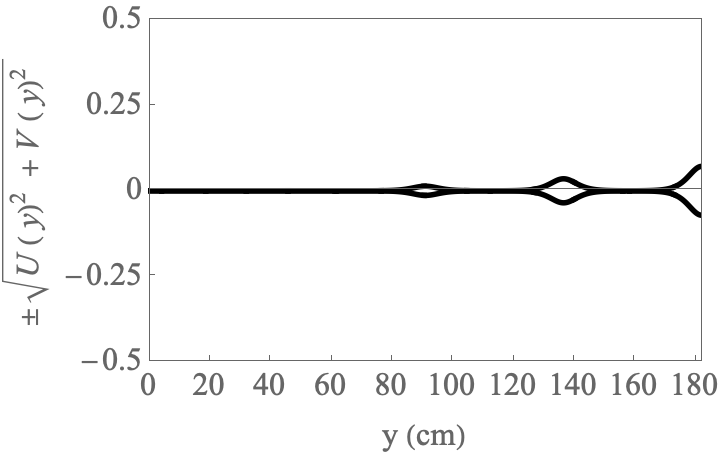}}
\caption{\label{figsnstab} Perturbation variation in $y$ for $n=4$.}
\end{figure}
The result shown in Figure~{\ref{figsnstab} has a left-right asymmetry, which the wavetank does not have. However, 
as discussed in \S{{\ref{experiments}}, surface displacement is measured with an in-situ probe that traverses the
tank from left to right. It is possible that the motion of the gage through the air-water interface introduces a
symmetry-breaking perturbation.

\section{Experimental Apparatus and Procedures}
\label{experiments}
The experimental apparatus comprised a wave basin, a wavemaker array, four wave gages, computer systems and water supply. 
It is described in detail in \cite{hhs05} with further procedures described in \cite{hps06}.
The wave basin was 12 ft long, 6 ft wide and 1 ft deep. 
Along one 6 ft endwall was a segmented wavemaker composed of 32 individually vertically oscillating triangular wedges of 2.25 in width.
The motion of each paddle was independently programmable and controlled by real-time computers using dual feedbacks from each paddle.
Above the basin was an $x, y, z$-positioning system for a wave gage array that could traverse the tank in a prescribed direction at a prescribed speed. The vertical dimension was used for wave-gage calibration. 

The Corian bottom and vertical glass sidewalls were cleaned with alcohol before the tank was filled with untreated tap water to a depth of $h > 20$ cm. A brass bar that spanned the width of the tank and was mounted on a moveable carriage above the tank skimmed the surface film to the end of the basin.
There, the film was vacuumed with a wet vac until the depth was $h = 20$ cm.  The tank was allowed to settle for a minimum of 10 min before each experiment, and each set of experiments was conducted within a
2 hr period after cleaning the surface.  This helped to reduce dissipative effects caused by the surface film.
The frequency, amplitude, and phase of the paddles were controlled to generate obliquely interacting wavetrains. To this end,
the paddles were programmed with a displacement, $\eta_p(y, t)$ given by
\beq
\eta_p(y, t) = a_p \cos\Big( \frac{n \pi}{W} y_j\Big) \cos(\omega t) + a_{3p} \cos\Big(\frac{3 n \pi}{W} y_j\Big) \cos(\omega t),
\label{paddle}
\eeq
where the subscript $j=1, 2, \dots, 32$ indicates the paddle number so that $y_j= 2.25 (j-1)$ cm is the $y$-distance to the edge of the $j$th paddle.
The paddle motion given by ({\ref{paddle}}) includes the third-harmonic term whose presence was explained in \S{\ref{theory-vnlse}}.
Its importance was recognized by Fuhrman \& Madsen \cite{fm06} and verified experimentally by \cite{hps06}. It is essential in generating bi-periodic patterns of waves with nearly permanent form.  

The values of $a_{p}$ and $a_{3p}$ are not the same values as $a_{0}$ and $a_{3}$ in ({\ref{sdvnlse}}) because there is a transfer of energy between the mechanical wavemaker and the water motion that is not 1:1 or necessarily linear. 
See {\cite{hps06} for a review of linear and nonlinear wavemaker theory, which models this transfer. 
Here, we chose a value of $a_p$ and then had to determine the best value of $a_{3p}$ to obtain a uniform pattern of waves.
To do that, we first generated waves using ({\ref{paddle}}) with $a_{3p}=0$. 
We measured a time series of the surface displacement at $x=150$ cm from the wavemaker, computed its Fourier transform, and set $a_0$ to be the Fourier amplitude of the component at the carrier wave frequency. 
We note, however that the 3rd-harmonic term has the same frequency as the carrier wave, so some of the energy there is due to the 3rd-harmonic term. 
Second, using ({\ref{eqnb13}}) and referencing ({\ref{3rd}}) and ({\ref{sdvnlse}}), we computed a value for 
$a_3 = a_0^3 k^2 |b_{13}|/2$. 
Third, to account for the wavemaker-to-water transfer process, we obtained a ``predicted" value of $a_{3p}$ to be $a_3 {\rm{sgn}}(b_{13})/a_0$. 
This predicted value of $a_{3p}$ assumes a linear transfer function between wavemaker and water motion. 
Fourth, to account for nonlinearity that arises in the first and third steps,
we conducted experiments with the chosen value of $a_p$ and variable values of $a_{3p}$ nearby the predicted value
until we found the value of $a_{3p}$ that generated a uniform-amplitude wavetrain. See Table~{\ref{tableparams}} for the values used.
\begin{table}[ht]
\begin{center}
\begin{tabular}{ccccccccccc}
\hline
$n$                    &     $a_p$ (cm)         &   $ a_{3p}$ (cm)     \\
\hline
4                        &              0.80         &     -0.10  \\
5                        &              0.60         &     -0.10   \\
6                        &              0.60         &     -0.10   \\
7                        &              0.60         &    -0.10  \\
8                        &              0.60         &     -0.05  \\
9                        &              0.60         &    -0.05   \\
10                      &              0.60         &    -0.10  \\
11                      &              0.60         &     -0.10    \\
\hline
\end{tabular}
\end{center}
\caption {Number of nodal lines ($n$) and forcing amplitudes ($a_p$ and $a_{3p}$).  For all experiments, $\omega = 8\pi$/sec and $W=12$ ft.}
\label{tableparams}
\end{table}

For all of our experiments, $f=4$Hz was the cyclic frequency so that $\omega=8\pi$/sec, and the wavenumber from ({\ref{dispersingle}}) was $\kappa = 0.626$/cm.
The value of $\kappa h =12.5$ so that the deep-water approximation is valid.
The relative strength of gravity vs surface tension is measured by the Bond number, $B_0 = g/ \sigma\kappa^2 = 36$ so that surface tension has a small
effect on the waves. Nevertheless, we included surface tension in the dispersion relation and the calculations of the coefficients of the sNLSE ({\ref{snlse}}). Comparisons of the amplitude of the Stokes-type solution of vNLSE ({\ref{vnlse}}) with measurements do not require the use of the coefficients in that equation. Comparisons of the amplitudes of the sn solution of sNLSE require the values of $\mu$ and 
$\chi$. These values as well as the $x$-- and $y$--wavenumbers are listed in Table~{\ref{tablecoefs}}. 
\begin{table}[ht]
\begin{center}
\begin{tabular}{ccccccccccc}
\hline
$n$                    &     $k_n$ (1/cm)         &   $ l_n$ (1/cm)    &   $\mu (cm^2/s)$   &   $\chi (cm^{-2}s^{-1} )$    \\
\hline
4                        &              0.622         &     0.069          &16.666 	&4.519  \\
5                        &              0.620         &     0.086         &  16.508	& 4.520  \\
6                        &              0.617         &     0.103         &  16.316 	& 4.522  \\
7                        &              0.614         &    0.120         &  16.089	  	&4.524 \\
8                        &              0.611         &     0.137         &15.827	 	& 4.526\\
9                        &              0.606         &    0.155         &  15.530 	&4.528  \\
10                      &              0.602         &    0.172         & 15.198  	&4.531  \\
11                      &              0.597         &     0.189         &   14.831 	& 4.534 \\
\hline
\end{tabular}
\end{center}
\caption {Number of nodal lines, the $x$-- and $y$--wavenumbers, and the values of the coefficients that 
are required to compute the sn solution ({\ref{exactsurface}}) with ({\ref{exact}}) of the sNLSE.
For all experiments, $\omega = 8\pi$/sec and $\kappa = 0.626$/cm.}
\label{tablecoefs}
\end{table}

Four wave-gages were supported along a line in the $x$--direction above the basin using the $x, y, z$-positioning system. 
The gages were of capacitance-type,
in-situ probes with a diameter of about 1 mm. For all experiments, except the one shown in Figure~{\ref{gendata}}b, the gages were positioned at $x_1=60$ cm, $x_2=90$ cm, $x_3=120$ cm, and
$x_4=150$ cm away from the wavepaddles and traversed the tank in the $y$--direction at a speed of $v_c = 5$ cm/s.
They provided time-series of surface displacement as the waves propagated across them in the $x$--direction. As they traversed the tank
in the $y$--direction, they were not measuring a single crest across the tank, but were measuring many crests that went across them
at the varying $y$-values at a fixed $x$. An example of the measured time series is shown in Figure~{\ref{gendata}}a for the $n=10$
experiment. Each group of oscillations corresponds to a half-period of a full envelope cycle. 
The number above each group refers to the number of the half-period, and is used in the discussion below of how we obtained the parameters required to compute the solutions.

We note that we did not observe modulational instability in the $x$--direction. For example, Figure~{\ref{gendata}}b shows a wave profile measured by a wave gage at a fixed value of $y$ that traversed the tank
in the $x$--direction. The gage traveled at a speed of 4 cm/s, so there is a Doppler shift. Therefore the wavelengths of the carrier wave in that plot are about a third of the actual $x$-wavelength.  The  amplitude variations at the end of the time-series, starting at $x \approx 210$ s, are the result of reflections from the tank endwall at $x=365$ cm. The experiments are stopped after about that time. The amplitudes of the waves do not show evidence of modulational instability.
\begin{figure}
\centerline{\includegraphics[width=2.5in]{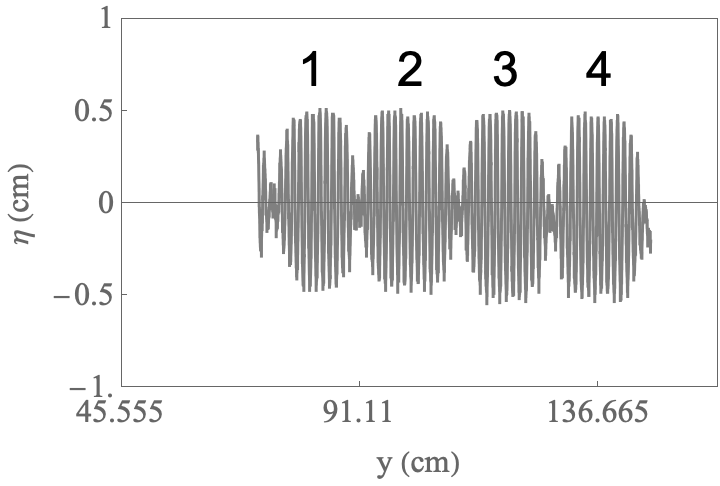}~~~~~~~~\includegraphics[width=2.5in]{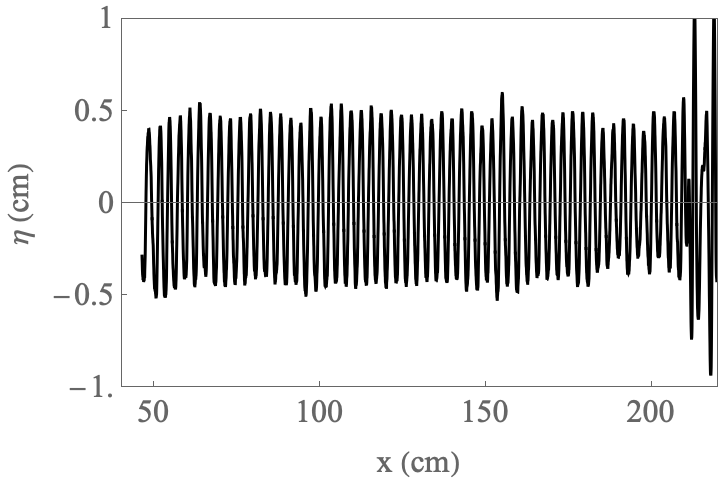}}
\hspace{4.5cm} (a) \hspace{6.5cm} (b)
\caption{\label{gendata} (a) Surface displacement in the transverse direction in the experiment with $n=10$ obtained at $x_3=120$cm. 
The numbers $1, 2, 3, 4$ correspond to the first through fourth half periods.
(b) Surface displacement as a function of propagation distance, $x$, at fixed $y=2 W/5 = 72.89$ cm
for the experiment with $n=5$ nodal lines.
The  amplitude variations at the end of the record, $x \approx 210$s, are the result of reflections from the tank endwall.}
\end{figure}



To analyze the data, we subtracted out the mean, and applied the calibration, which was obtained by moving the gages vertically in the positive and negative directions at fixed $x$ and $y$ values.  
To fit the Stokes-type solution of the vNLSE and the sn solution of the sNLSE we used the following procedure:
   \begin{enumerate}
   \label{procedure}
   \item  We measured the amplitude of each individual crest in an integer number of half-periods of the amplitude envelopes in the
$y$--direction for each experiment.
The number of half-periods for experiments with $n=4, 5$ was 1; for  experiments with $n=6, 7, 8$ was 2, for $n=9$ was 3; and for 
$n=10, 11$ was 4.
   \item We fit a curve to those points using the interpolation function of Mathematica.
   \item We digitized the fitted curve with $N=500$ points. Call those points $a_i^{experiment}$, $i=1, 2, ..., N$. 
   \item We computed the theoretical solution and digitized the same number of half-periods of the theoretical solution with $N$ points.
         \begin{enumerate}
         \item For the Stokes-type solution of the vNLSE, we chose an amplitude, $a_0$, computed the corresponding
         value of $a_3$ from ({\ref{a3}}) and
         computed the theoretical solution for the surface displacement from ({\ref{sdvnlse}}).
         \item For the sn solution of the sNLSE, we chose a value for $m$, calculated the corresponding
         values of $c$ and $a_{sn}$ from ({\ref{a0c}}), and computed the theoretical solution for the surface displacement from
         ({\ref{exactsurface}}).
         \end{enumerate}
            For either case, call the points digitized from the theoretical solution for the surface displacement $a_i^{predicted}$, $i=1, 2, ..., N$.
     \item We computed an error, $\scrE$, between the predicted and measured values given by
\beq
\scrE = \frac{  \sum_{i=1}^N  \Big(  | a_i^{experiment} | - |a_i^{predicted}| \Big)^2}{  \sum_{i=1}^N   | a_i^{experiment} | ^2},
\label{eqnerror}
\eeq
($N=500$)
for the data at each of the four wave gages. Then we took the average of those four values.
       \item We iterated this procedure until we minimized the average value of the errors.     We used the values of $a_0$ or $m$ that gave this minimized average error to produce the corresponding theoretical predictions. Then the theoretical prediction used in comparisons      in \S{\ref{results}} is the same for each wave gage location.
\end{enumerate}

\section{Results}
\label{results}
In this section we present measurements of the amplitude variation in the $y$--direction of bi-periodic wavetrains and determine
how well the solutions of vNLS and sNLS describe them.  We also consider the extra variation of amplitudes that are observed
in the $y$-direction.
%

\subsection{Envelope from the vNLSE}
\label{resultsvnlse}

Figure~{\ref{figdatav}} shows the measured surface displacement (the gray curves) in the transverse direction, $W/4 < y < 7 W/8$, from the four gages for each of the seven experiments (except for the $n=7$ experiment, for which we did not have data from the first two gages). The oscillations of the carrier wave were fast enough that the measured curves are very close together.
The envelope curves are given by ({\ref{sdvnlse}}), with $x$ and $t$ set to zero to obtain the $y$-variation without the fast oscillations. 
The values of $a_0$ and $a_3$ are determined by the procedure outlined at the end of \S{\ref{experiments}} and are listed in
Table~{\ref{tablevexact}}.
\begin{table}[ht]
\begin{center}
\begin{tabular}{ccccccccccc}
\hline
$n$                    &     $a_0$ (cm)      &             $a_3$ (cm)     \\
\hline
4                        &          0.521     &                   -0.089  \\
5                        &          0.574     &                   -0.049        \\
6                        &          0.520     &                   -0.021     \\
7                        &          0.538     &                   -0.016 \\
8                        &           0.535    &                   -0.011 \\
9                        &           0.534    &                   -0.008           \\
10                      &           0.564    &                   -0.008           \\
11                      &           0.583    &                  -0.007   \\
\hline
\end{tabular}
\end{center}
\caption {Amplitudes, $a_0$ and $a_3$, used to compare the Stokes-type solution of vNLSE,
({\ref{sdvnlse}}) for $x=t=0$, with measurements.
}
\label{tablevexact}
\end{table}

An estimate of the error, $\scrE$, between the measured envelope of the $y$--profile and the envelope predicted by the Stokes-type solution of vNLSE, as given in ({\ref{sdvnlse}}) with values listed in Table~{\ref{tablevexact}} and with $x=t=0$, is given by ({\ref{eqnerror}}) using the procedure outlined in \S{\ref{experiments}}. 
The errors are listed in Table~{\ref{tableerror}} in the second through fifth columns for the time series from the four gage sites, $x_i,~ i=1, ..., 4$, of the experiments that varied the number of nodal lines from $n=4, 5, ... 11$ with and without $a_3$.
The sixth through ninth columns list the errors for the sn solution of sNLSE and are discussed in \S{\ref{resultssnlse}}. 
The last column of Table~{\ref{tableerror}} lists $\epsilon_y = l_n/\kappa$, the measure of two-dimensionality of the wave patterns.
The following are some observations from the results listed in the Table:
  \begin{enumerate}
  \item Even though there is no constraint on the measure of two-dimensionality of the wave patterns, $\epsilon_y$, for the Stokes-type solution of vNLSE, the error is larger for experiments with the larger values of $\epsilon_y$. 
  \item The error when the third-harmonic term is not included (i.e.~when $a_3=0$) is significantly larger than when it is included 
  ($a_3 \ne 0$) for $n=4, 5, 6, 7$ ($0.110 \le \epsilon_y \le 0.192$).
    \item The error appears independent of the inclusion of the third-harmonic term for experiments with
     $n=8, 9, 10, 11$ ($0.220 \le \epsilon_y \le 0.302$).
    \item  For experiments with $n=4, 5, ...8$, ($0.110 \le \epsilon_y \le  0.220$), the error is less than 5\% if the third-harmonic term is included. 
   \item The maximum error is about 10\%.
   \end{enumerate}

In general, the Stokes-type solutions of the vNLSE that describes the amplitude variation in $y$ are in reasonable agreement with the data.
The inclusion of the third-harmonic term decreases the error substantially for the lower values of $\epsilon_y$. The improvement is visually apparent in Figure~{\ref{tablevexact}}; for $n=4, 5$ especially; one can see that the amplitude variation in
$y$ is
not described by a single sinusoidal mode. 

\begin{figure}

{\tiny{$n=4$}}\\

\vspace{-1.5cm}
\hspace{3cm}(a) \hspace{3.2cm} (b) \hspace{3.2cm} (c) \hspace{3.2cm} (d)

~~~~~~~~~~~~\includegraphics[width=1.5in]{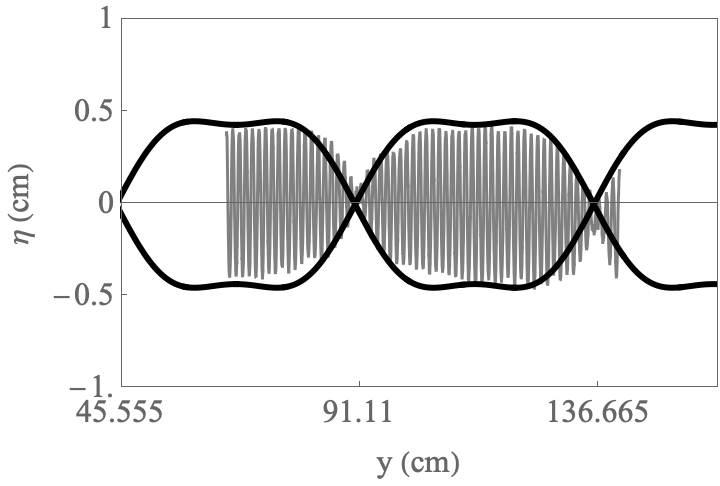}~
\includegraphics[width=1.5in]{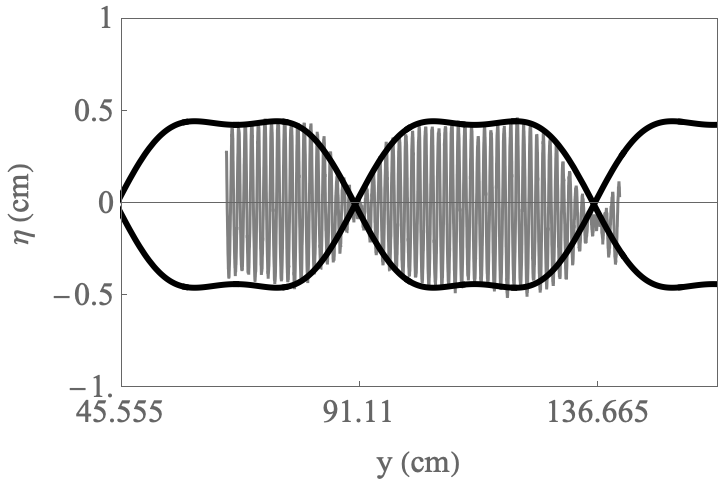}~
\includegraphics[width=1.5in]{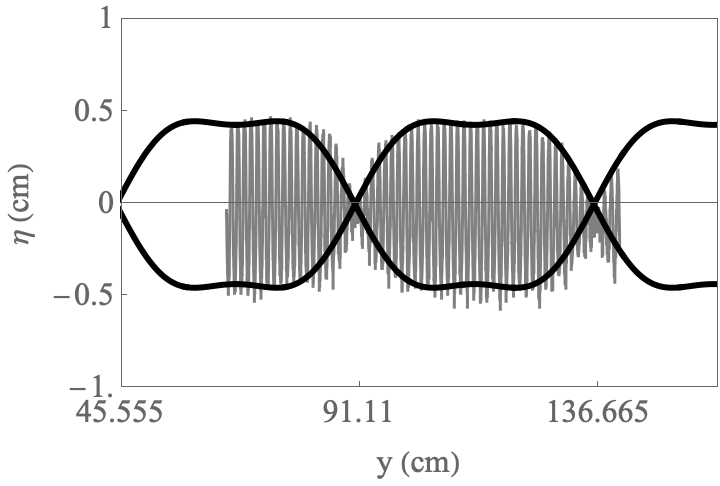}~
\includegraphics[width=1.5in]{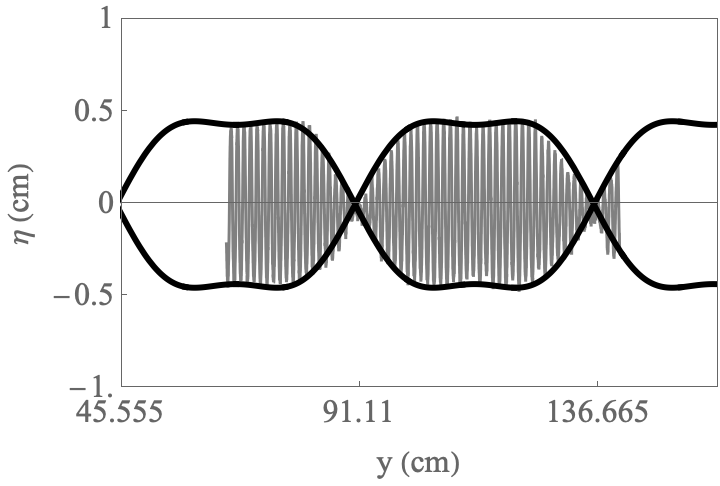}\\
\vspace{-0.2cm}
{\tiny{$n=5$}}\\

\vspace{-1.5cm}
~~~~~~~~~~~~\includegraphics[width=1.5in]{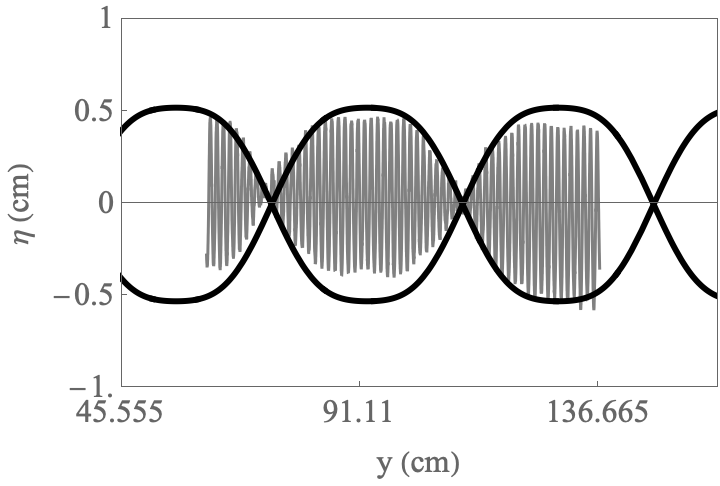}~
\includegraphics[width=1.5in]{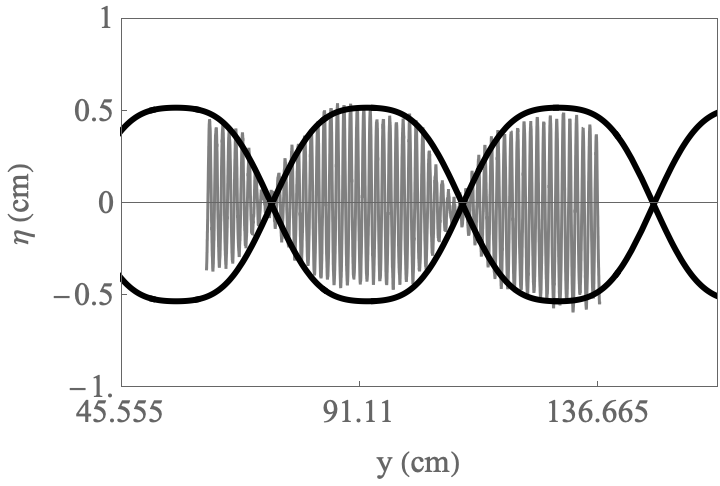}~
\includegraphics[width=1.5in]{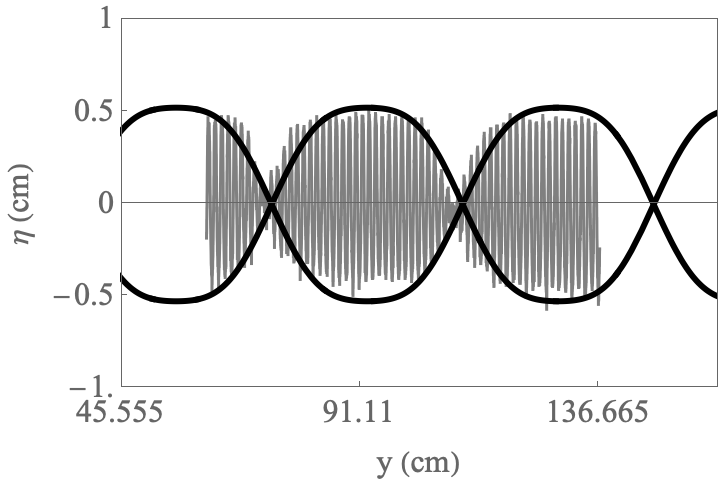}~
\includegraphics[width=1.5in]{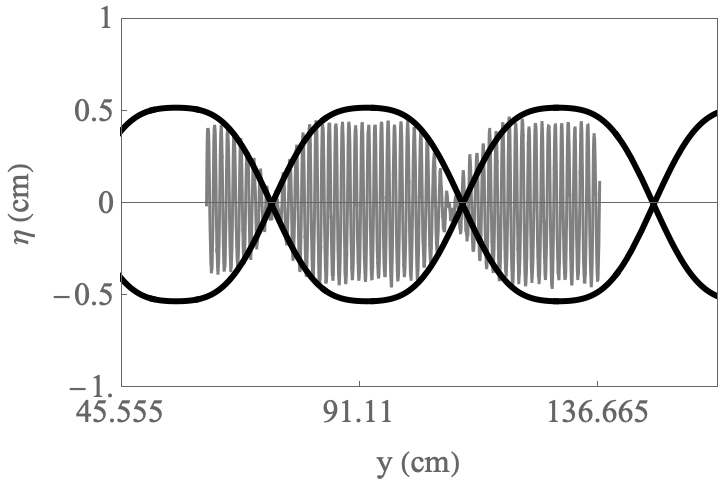}\\
\vspace{-0.2cm}
{\tiny{$n=6$}}\\

\vspace{-1.5cm}
~~~~~~~~~~~~\includegraphics[width=1.5in]{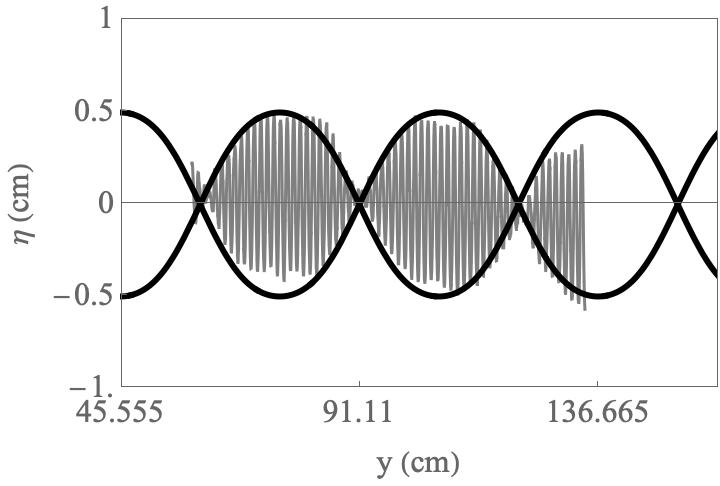}~
\includegraphics[width=1.5in]{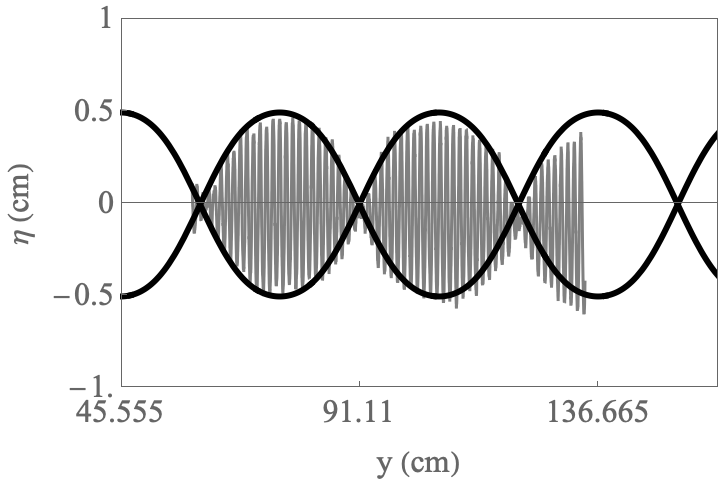}~
\includegraphics[width=1.5in]{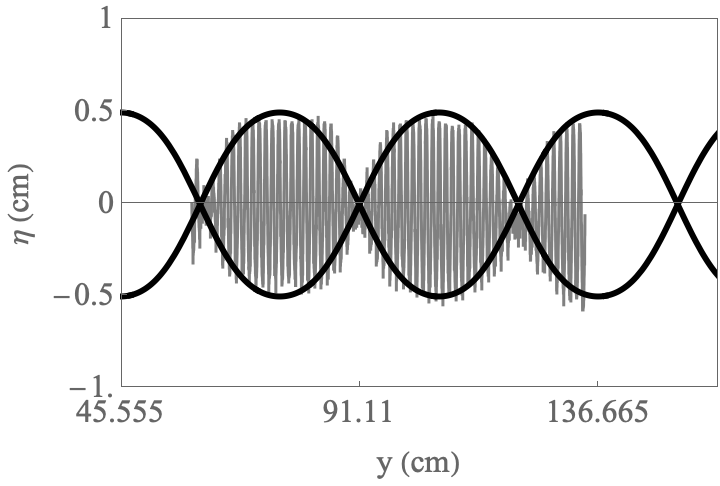}~
\includegraphics[width=1.5in]{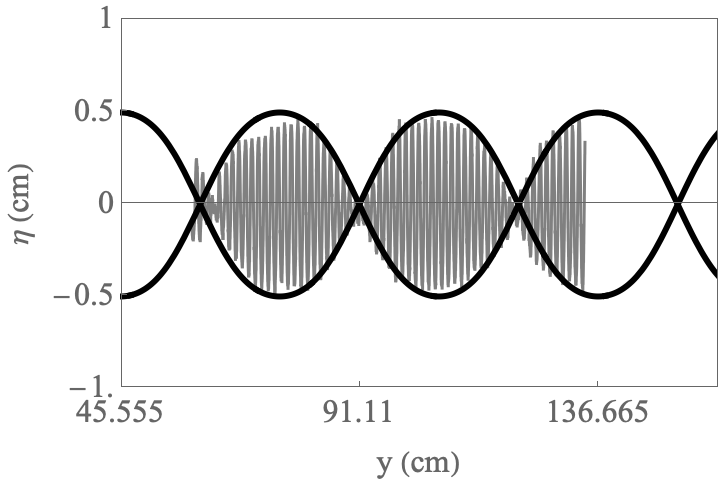}\\
\vspace{-0.2cm}
{\tiny{$n=7$}}\\

\vspace{-1.5cm}
~~~~~~~~~~~~\includegraphics[width=1.5in]{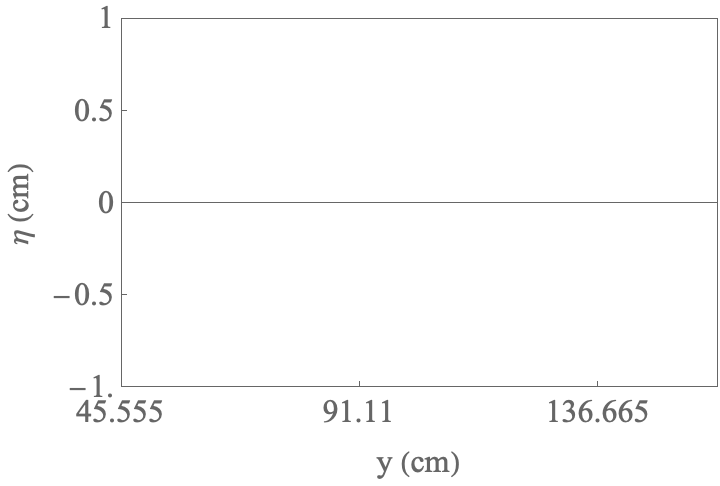}~
\includegraphics[width=1.5in]{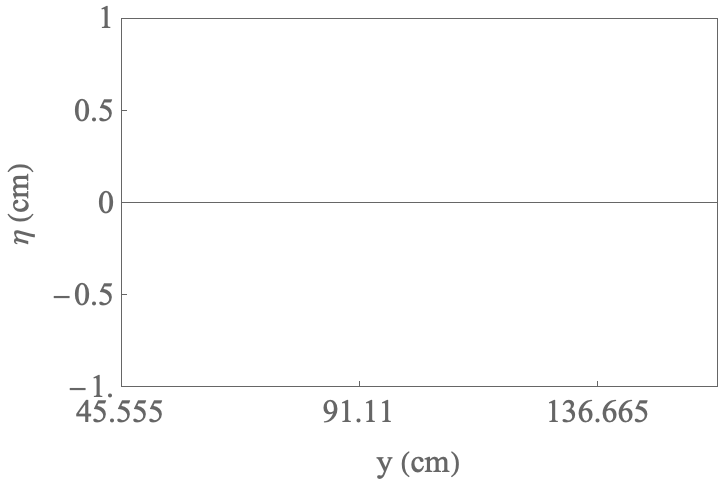}~
\includegraphics[width=1.5in]{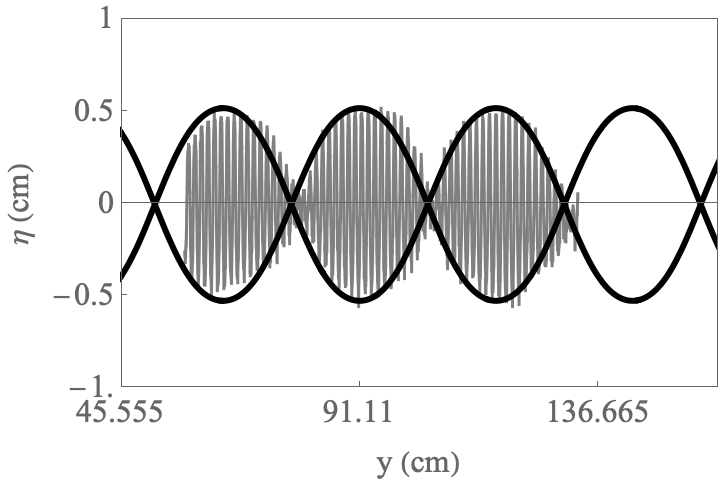}~
\includegraphics[width=1.5in]{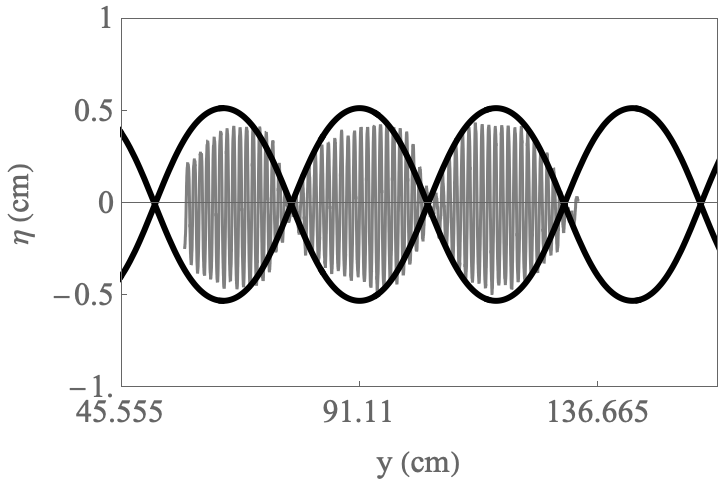}\\
\vspace{-0.2cm}
{\tiny{$n=8$}}\\

\vspace{-1.5cm}
~~~~~~~~~~~~\includegraphics[width=1.5in]{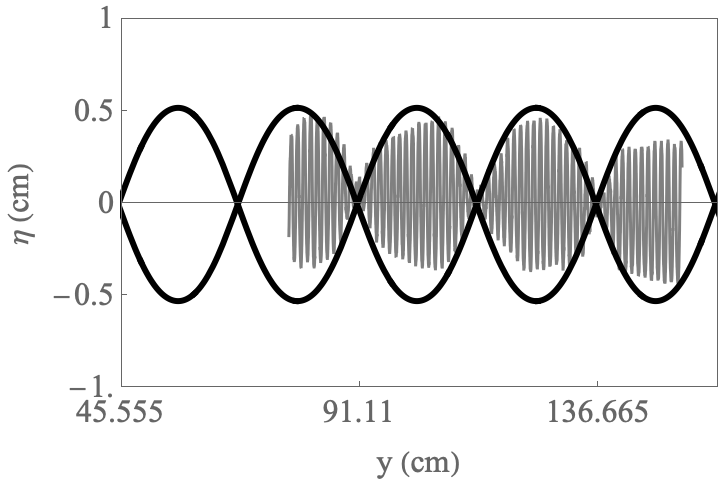}~
\includegraphics[width=1.5in]{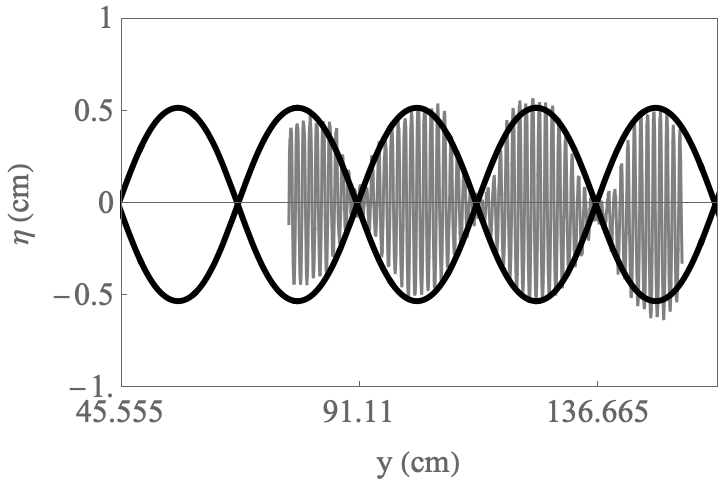}~
\includegraphics[width=1.5in]{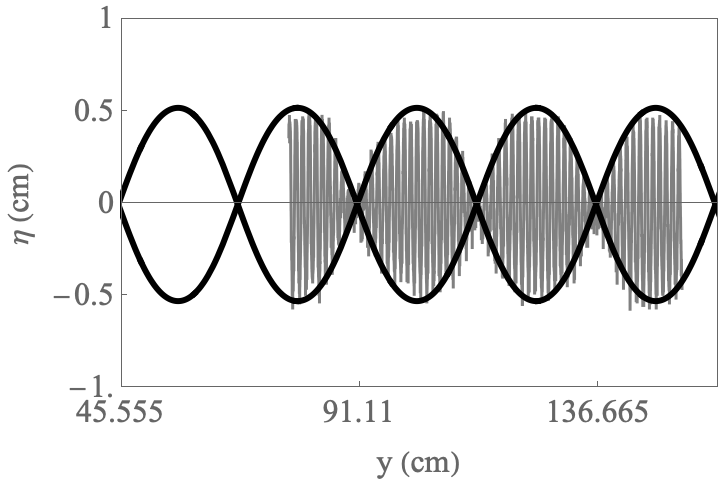}~
\includegraphics[width=1.5in]{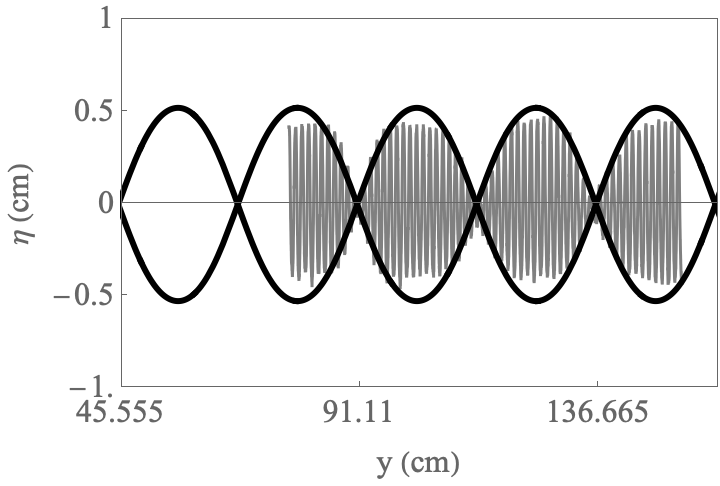}\\
\vspace{-0.2cm}
{\tiny{$n=9$}}\\

\vspace{-1.5cm}
~~~~~~~~~~~~\includegraphics[width=1.5in]{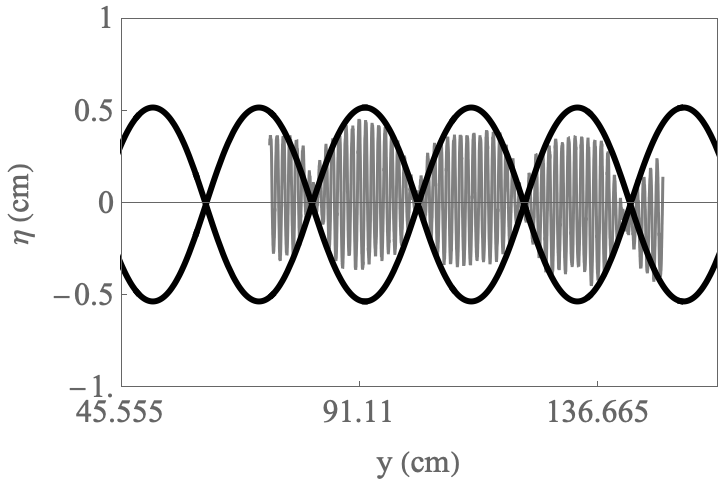}~
\includegraphics[width=1.5in]{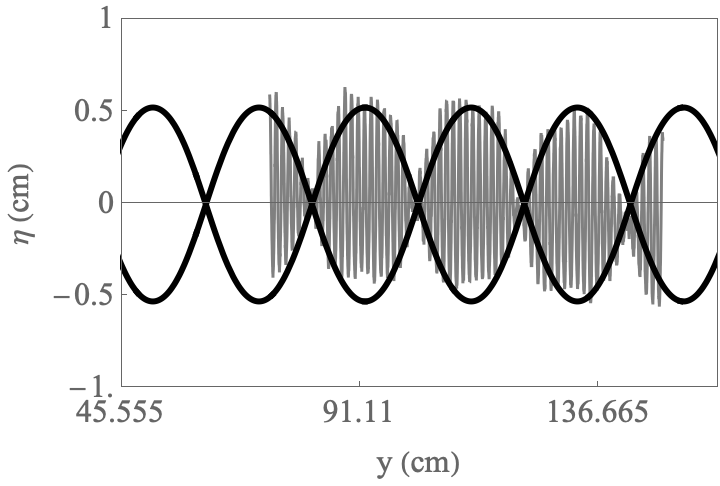}~
\includegraphics[width=1.5in]{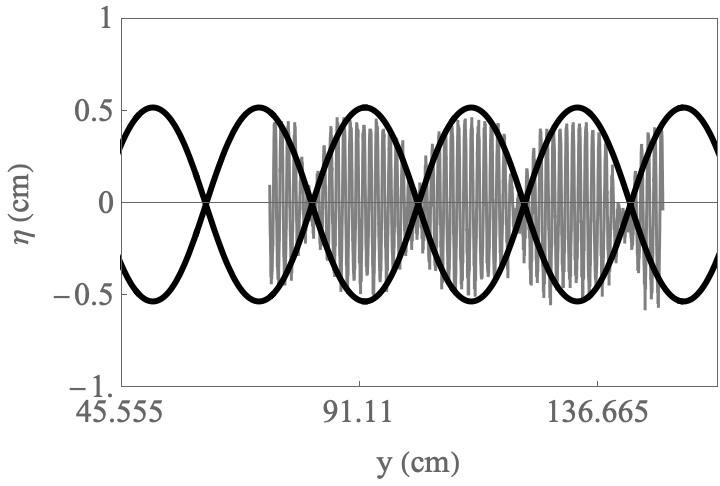}~
\includegraphics[width=1.5in]{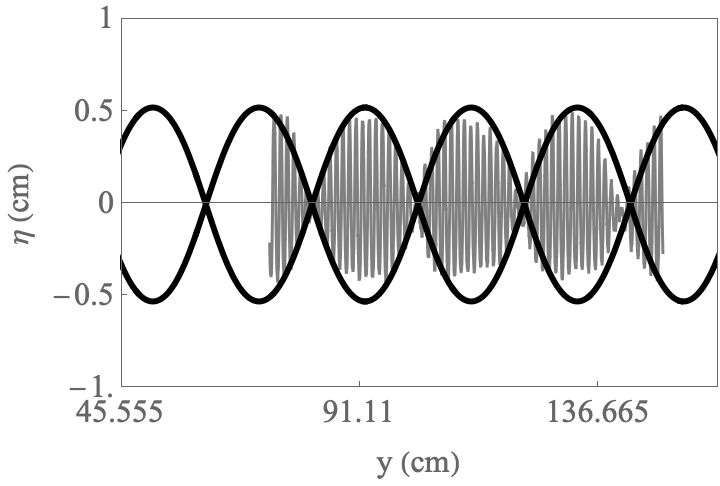}\\
\vspace{-0.2cm}
{\tiny{$n=10$}}\\

\vspace{-1.5cm}
~~~~~~~~~~~~\includegraphics[width=1.5in]{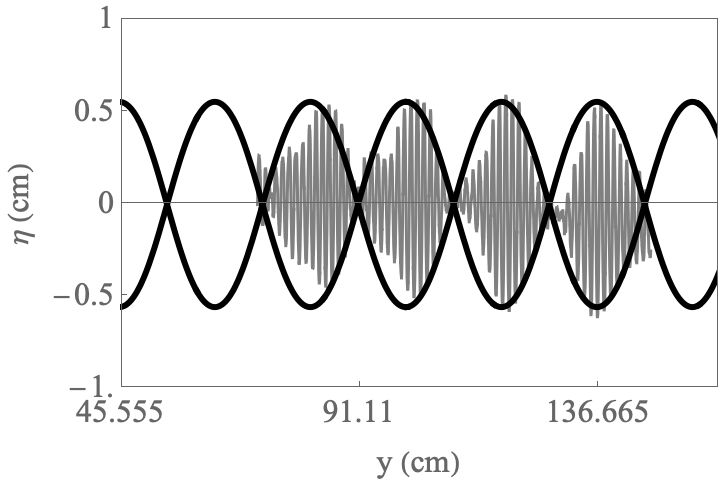}~
\includegraphics[width=1.5in]{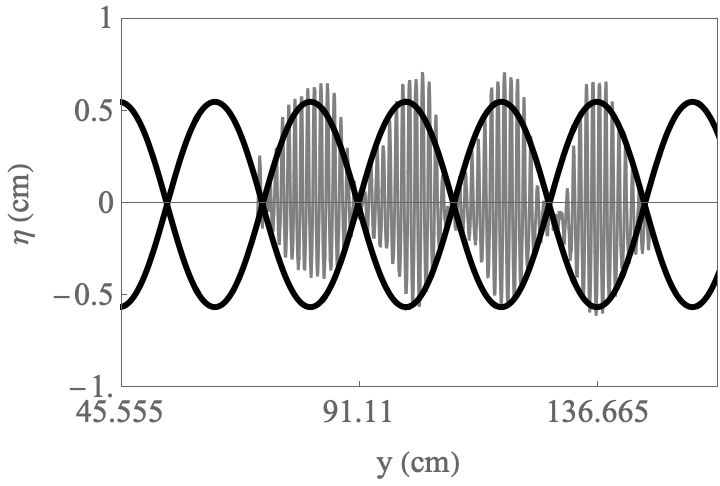}~
\includegraphics[width=1.5in]{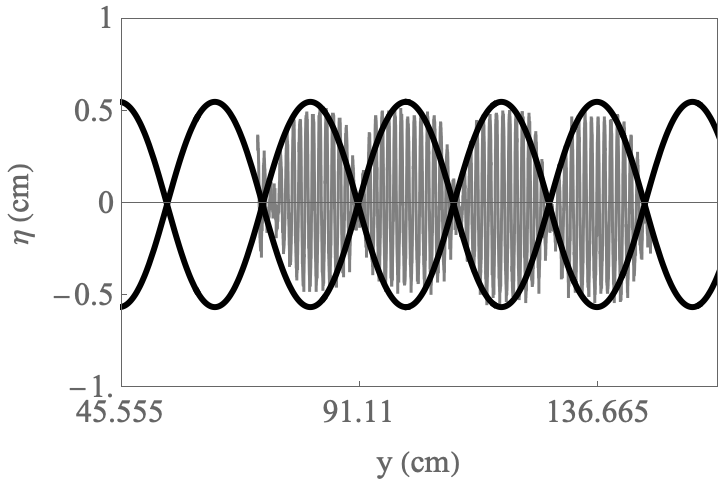}~
\includegraphics[width=1.5in]{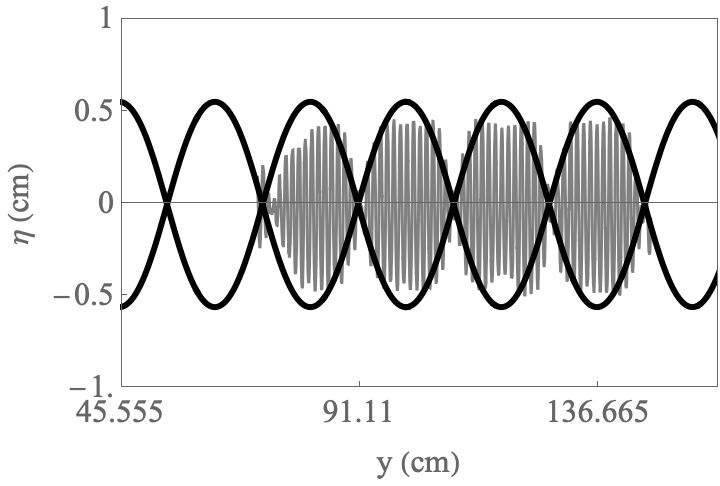}\\
\vspace{-0.2cm}
{\tiny{$n=11$}}\\

\vspace{-1.5cm}
~~~~~~~~~~~~\includegraphics[width=1.5in]{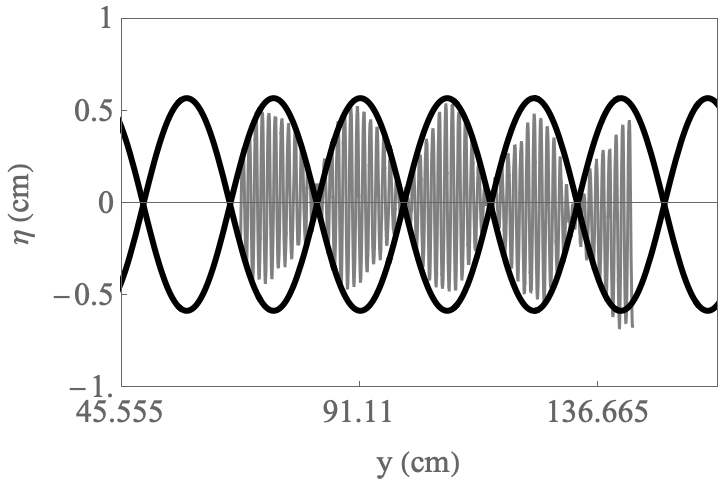}~
\includegraphics[width=1.5in]{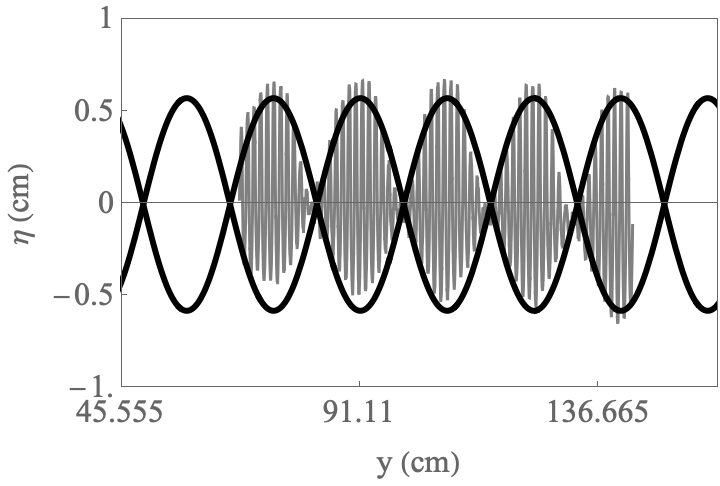}~
\includegraphics[width=1.5in]{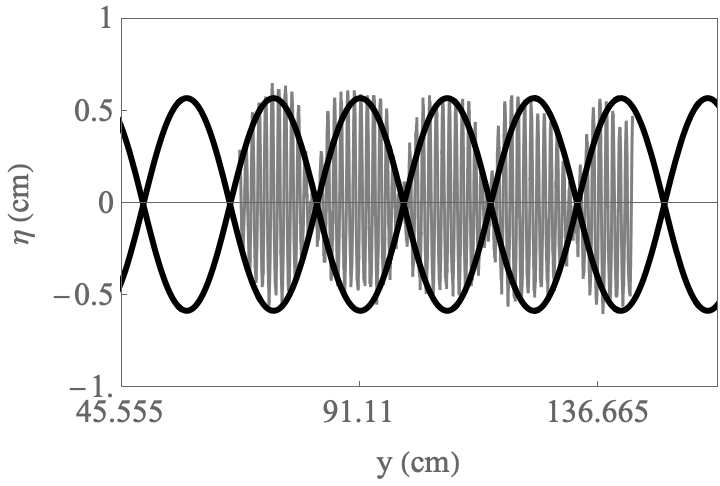}~
\includegraphics[width=1.5in]{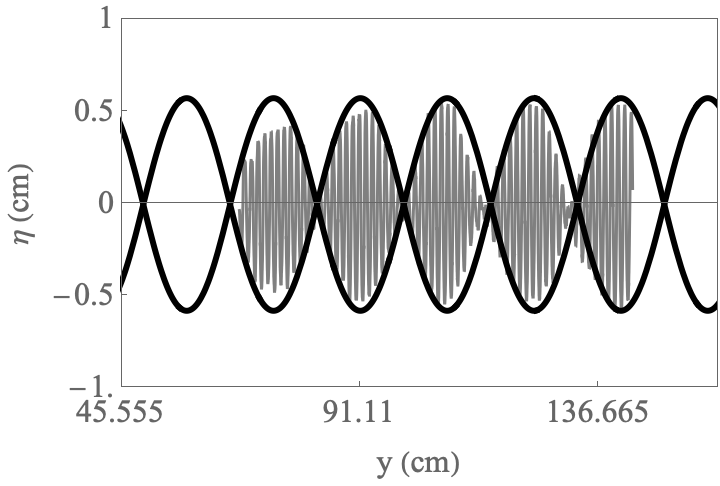}
\caption{\label{figdatav} Surface displacement in the transverse direction (gray curves) in experiments with $n=4, 5, \dots, 11$ in each row as indicated for (a) $x_1=60$ cm, (b) $x_2=90$ cm, (c) $x_3=120$ cm, (d) $x_4=150$ cm from the wave paddles. 
The envelope (black) curves are from the Stokes-type solution of the vNLSE given by ({\ref{sdvnlse}}) with $x=t=0$.
Parameters for the vNLSE solutions are given in Table~{\ref{tablevexact}}.}
\end{figure}


\begin{table}[ht]
\begin{center}
\begin{tabular}{| l || l c c c  c | l c  c  c  c | |c | c  |c | }
\hline
$n$                    &       &&    vNLSE    & &  &                    &     &&      sNLSE       & &   $\epsilon_y ={l_n}/{\kappa}$   \\
                        &   ~~$x_1$ & ~~$x_2$ &  $x_3$ &  $x_4$ &      & &   $x_1$ &  ~~$x_2$ &  $x_3$ &  $x_4$  &             \\
\hline
4                        &   0.012 & ~~ 0.004 &  0.009 &  0.006  &      & &   0.011 &  ~~0.007 &  0.010 &  0.005  &    0.110          \\
~~ (w/ $a_3=0$)&   0.040 & ~~ 0.036 &  0.044 &  0.020  &      & &         ~ &  ~ & ~ &  ~  &             \\
\hline
5                        &   0.039& ~~ 0.026 &  0.049 &  0.037  &      & &   0.011 &  ~~0.006 &  0.018 &  0.012  &   0.137          \\
~~ (w/ $a_3=0$)&   0.060 & ~~ 0.043 &  0.078 &  0.068  &      & &         ~ &  ~ & ~ &  ~  &             \\
\hline
6                        &   0.007 & ~~ 0.017 &  0.016 &  0.007  &      & &   0.005 &  ~~0.011 &  0.010 &  0.003  &   0.165          \\
~~ (w/ $a_3=0$)&   0.016 & ~~ 0.032 &  0.032 &  0.019  &      & &         ~ &  ~ & ~ &  ~  &             \\
\hline
7                        &   n/a     & ~~ n/a     &  0.027 &  0.039  &      & &   n/a    &  ~~n/a     &  0.018 &  0.029  &    0.192         \\
~~ (w/ $a_3=0$)&   n/a     & ~~ n/a     &  0.032 &  0.044  &      & &         ~ &  ~ & ~ &  ~  &             \\
\hline
8                        &   0.039 & ~~ 0.016 &  0.029 &  0.022  &      & &   0.025 &  ~~0.023 &  0.014 &  0.012  &     0.220        \\
~~ (w/ $a_3=0$)&   0.039 & ~~ 0.017 &  0.028 &  0.022  &      & &         ~ &  ~ & ~ &  ~  &             \\
\hline
9                        &   0.084 & ~~ 0.100 &  0.068 &  0.044  &      & &   0.048 &  ~~0.053 &  0.033 &  0.012  &    0.247         \\
~~ (w/ $a_3=0$)&   0.103 & ~~ 0.117 &  0.086 &  0.064  &      & &         ~ &  ~ & ~ &  ~  &             \\
\hline
10                      &   0.062 & ~~ 0.053&  0.048  &  0.066  &      & &   0.068 &  ~~0.64 &  0.049 &  0.064  &     0.274        \\
~~ (w/ $a_3=0$)&   0.061 & ~~ 0.057 &  0.058 &  0.076  &      & &         ~ &  ~ & ~ &  ~  &             \\
\hline
11                      &   0.038 & ~~ 0.032  &  0.059  & 0.012   &      & &   0.036  &  ~~0.039 &0.050   &0.016   &  0.302         \\
~~ (w/ $a_3=0$)&   0.048 & ~~0.032  & 0.070 &  0.010  &      & &         ~ &  ~ & ~ &  ~  &             \\
\hline 
\end{tabular}
\end{center}
\caption {Error, $\scrE$, computed using ({\ref{eqnerror}}) in comparisons of the predictions of amplitude envelope by ({\ref{sdvnlse}})
from the vNLSE and by ({\ref{exactsurface}}) from the sNLSE with $x=t=0$ and measurements obtained at the gage locations, $x_i$, $i=1, ... 4$, of the seven experiments. The
comparison  based on the solution of vNLSE is shown with ($a_3 \ne 0$) and without ($a_3=0$) the inclusion of the third-harmonic-in-$y$
term. The last column gives the measure of two-dimensionality of the wave patterns.}
\label{tableerror}
\end{table}

\subsection{Envelope from the sNLSE}
\label{resultssnlse}

Figure~{\ref{tablesnfit}} shows the measured surface displacement (the gray curves) in the transverse direction, $W/4 < y < 7 W/8$, from the four gages for each of the seven experiments (except for the $n=7$ experiment, for which we did not have data from the first two gages). The oscillations of the carrier wave were fast enough that the 
measured curves are very close together.
The envelope curves are from the sn solution of the sNLSE given by ({\ref{exactsurface}}), with $x$ and $t$ set to zero to obtain the $y$-variation without the fast oscillations. 
The values of the elliptic modulus, $m$, the amplitude, $a_{sn}$, and transverse wavenumber, $c$, used for the sn solutions
are determined by the procedure outlined at the end of \S{\ref{experiments}} and are listed in
Table~{\ref{tablesnfit}}.

The errors between the predictions of the amplitude variation in the $y$--direction from the sn solution of the sNLSE and the measured envelope were computed using ({\ref{eqnerror}}) and are shown in the sixth through ninth columns of Table~{\ref{tableerror}}.  
The following are some observations from the results listed in the Table:
    \begin{enumerate}
    \item The errors increase with increasing $\epsilon_y$. The increase is consistent with the approximation inherent in
    the sn solution of the sNLSE that $\epsilon_y \ll 1$. 
    \item For experiments with $n=4, 5, 6$ ($0.110 \le \epsilon_y \le 0.165$), the errors are about 1\% or less. 
    \item For all of the other experiments, except for $n=10$, the errors are less than 5\%; for the experiment with $n=10$, the error was around 5-7\%. 
    \end{enumerate}

In general, the sn solution of the sNLSE that describes the amplitude variation in $y$ are in good agreement with the data. 
The data show that the amplitude variation in the $y$--direction is not sinusoidal; there is a flattening of the $y$--envelope. The elliptic modulus 
of the sn function allows for this flattening to be modeled without requiring the addition of higher-order (in nonlinearity) terms. 

\begin{table}[ht]
\begin{center}
\begin{tabular}{ccccccccccc}
\hline
$n$                    &             $m$            &                $a_{sn}$ (cm)           &       $c$ (1/cm)  \\
\hline
4                        &              0.9696         &                    0.461                    &      0.124    \\
5                        &              0.9500         &                    0.516                    &      0.142    \\
6                        &              0.8763         &                    0.480                    &      0.144    \\
7                        &              0.8390         &                    0.506                    &      0.160    \\
8                        &              0.7853         &                    0.507                   &       0.173    \\
9                        &              0.7489         &                    0.523                   &       0.189       \\
10                      &              0.7184         &                    0.539                    &     0.205       \\
11                      &              0.6965         &                    0.560                   &      0.222   \\
\hline
\end{tabular}
\end{center}
\caption {Values of the elliptic modulus, $m$, the amplitude, $a_{sn}$, and transverse wavenumber, $c$, used for the sn solution for the seven different experiments.}
\label{tablesnfit}
\end{table}


\begin{figure}
{\tiny{$n=4$}}\\

\vspace{-1.5cm}
\hspace{3cm}(a) \hspace{3.2cm} (b) \hspace{3.2cm} (c) \hspace{3.2cm} (d)

~~~~~~~~~~~~\includegraphics[width=1.5in]{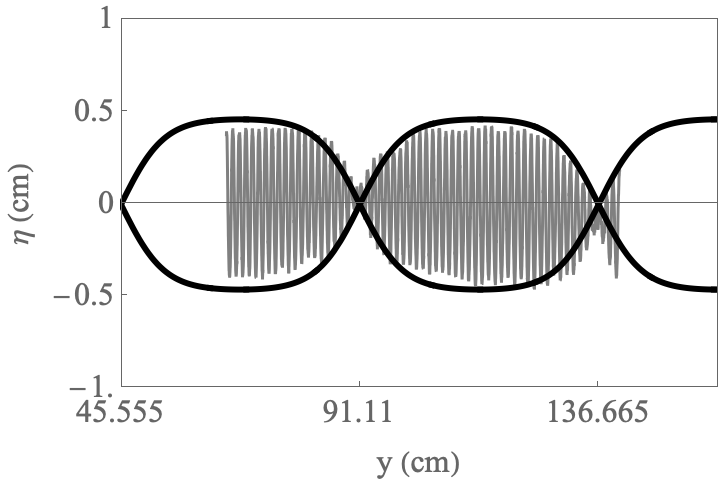}~
\includegraphics[width=1.5in]{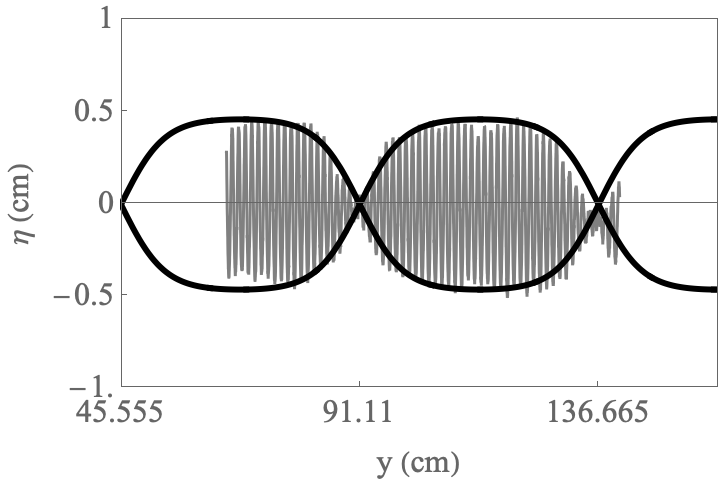}~
\includegraphics[width=1.5in]{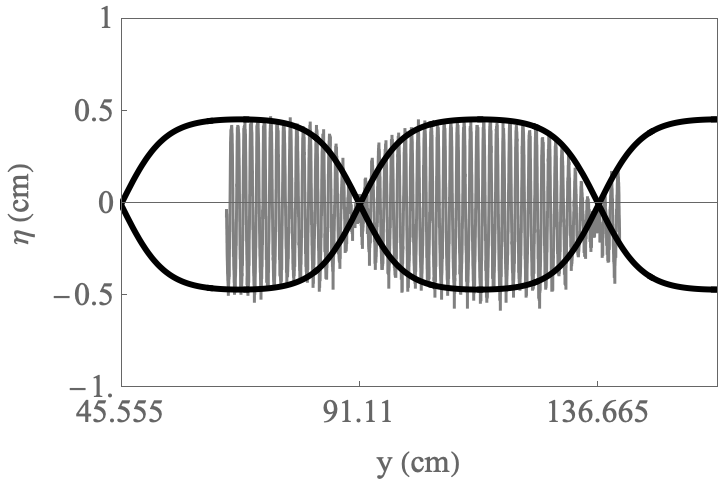}~
\includegraphics[width=1.5in]{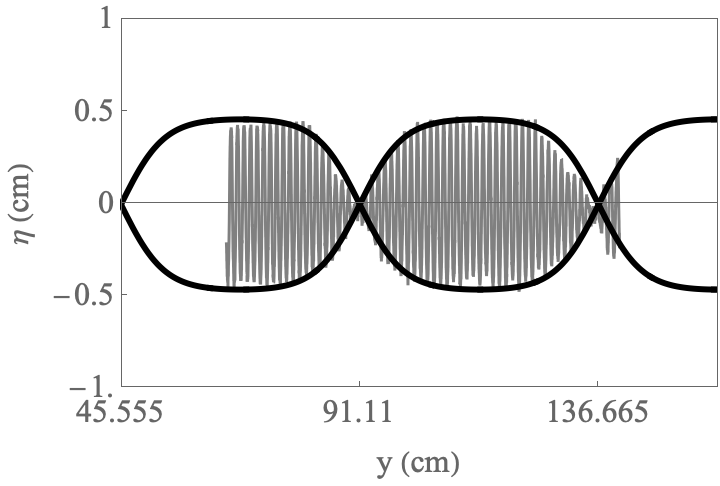}\\
\vspace{-0.2cm}
{\tiny{$n=5$}}\\

\vspace{-1.5cm}
~~~~~~~~~~~~\includegraphics[width=1.5in]{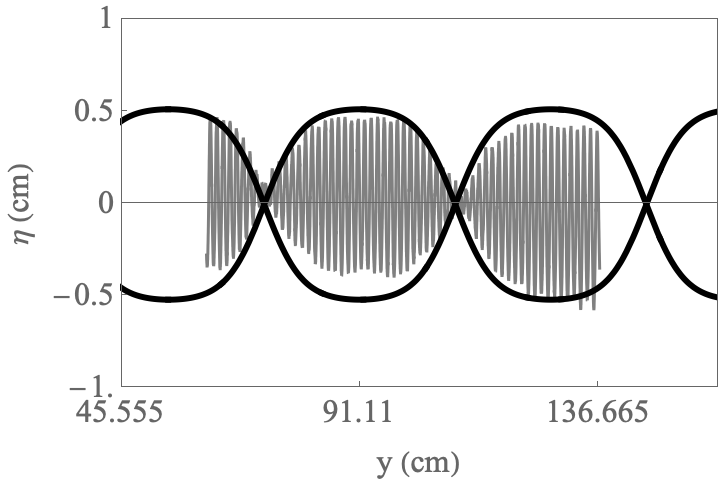}~
\includegraphics[width=1.5in]{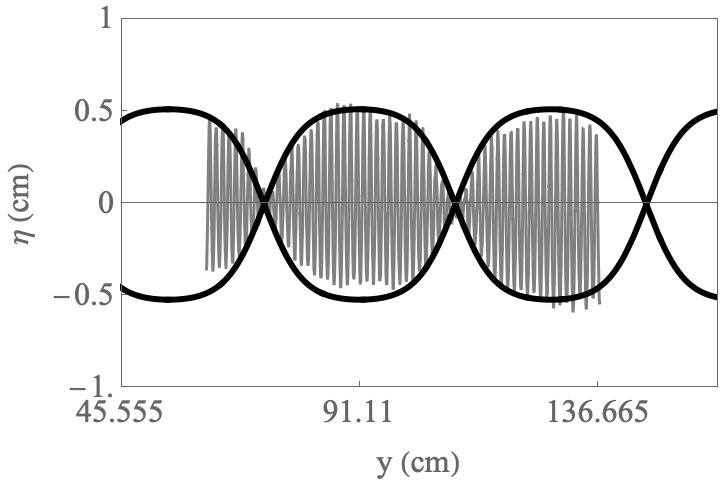}~
\includegraphics[width=1.5in]{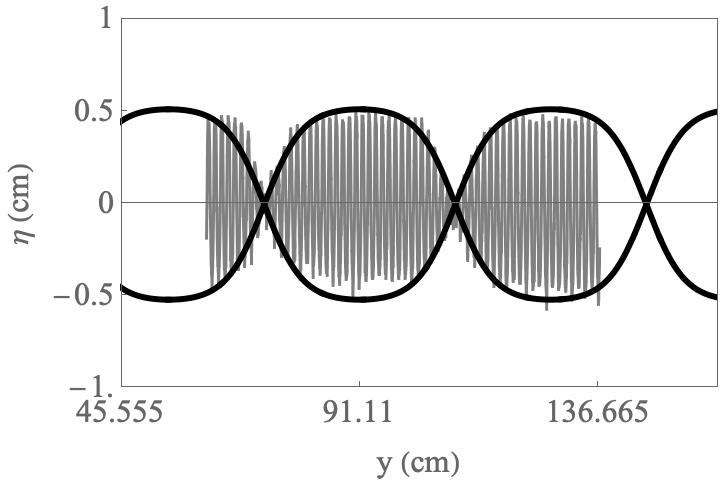}~
\includegraphics[width=1.5in]{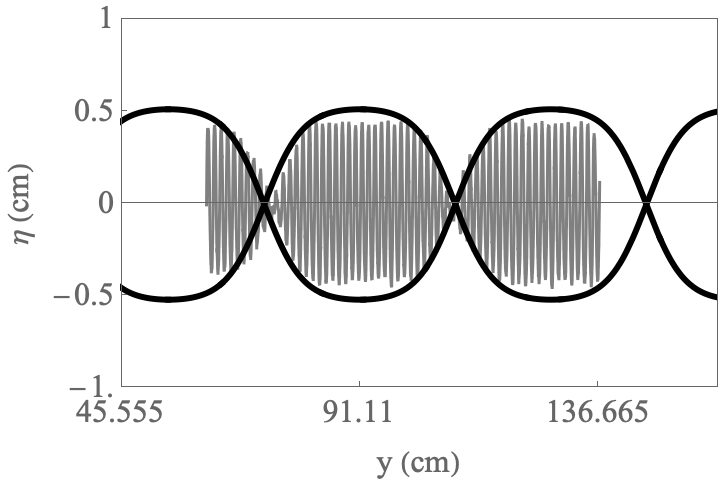}\\
\vspace{-0.2cm}
{\tiny{$n=6$}}\\

\vspace{-1.5cm}
~~~~~~~~~~~~\includegraphics[width=1.5in]{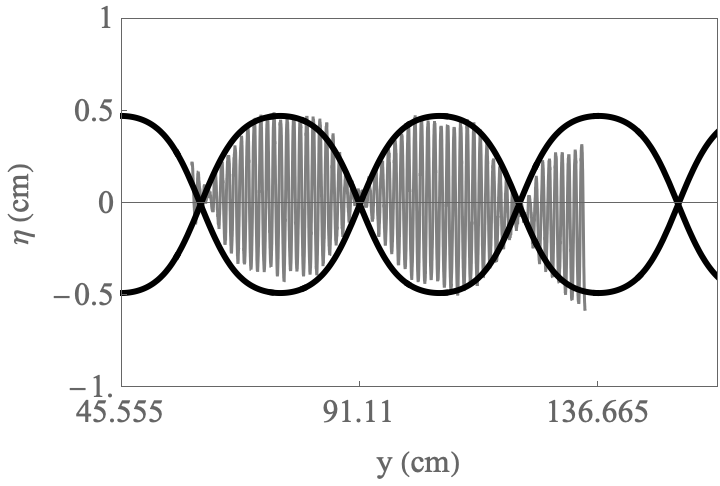}~
\includegraphics[width=1.5in]{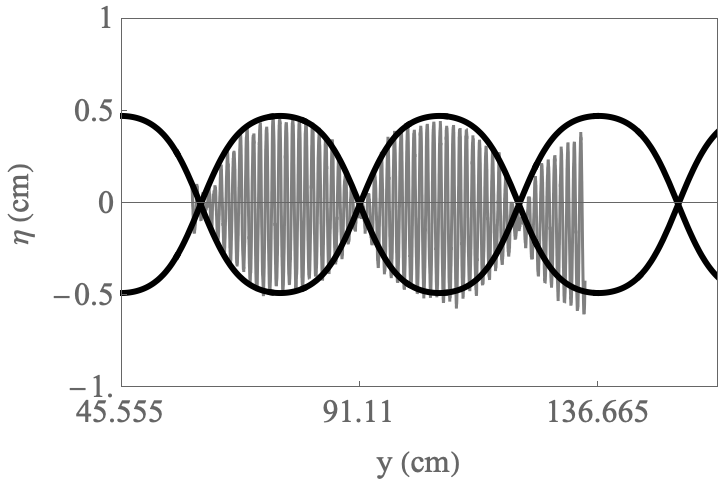}~
\includegraphics[width=1.5in]{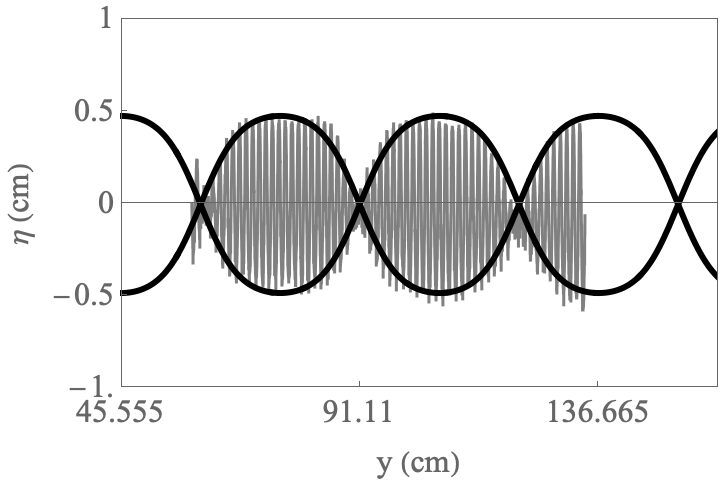}~
\includegraphics[width=1.5in]{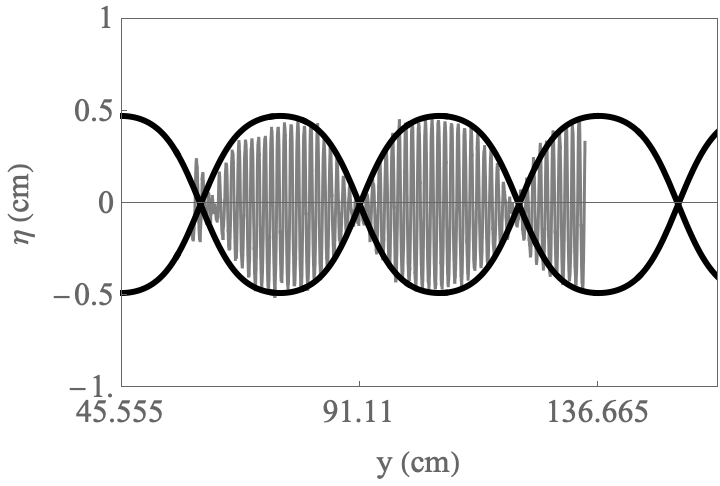}\\
\vspace{-0.2cm}
{\tiny{$n=7$}}\\

\vspace{-1.5cm}
~~~~~~~~~~~~\includegraphics[width=1.5in]{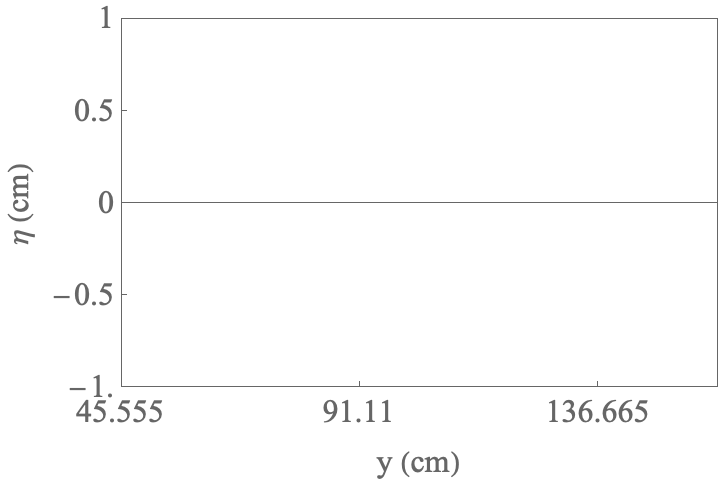}~
\includegraphics[width=1.5in]{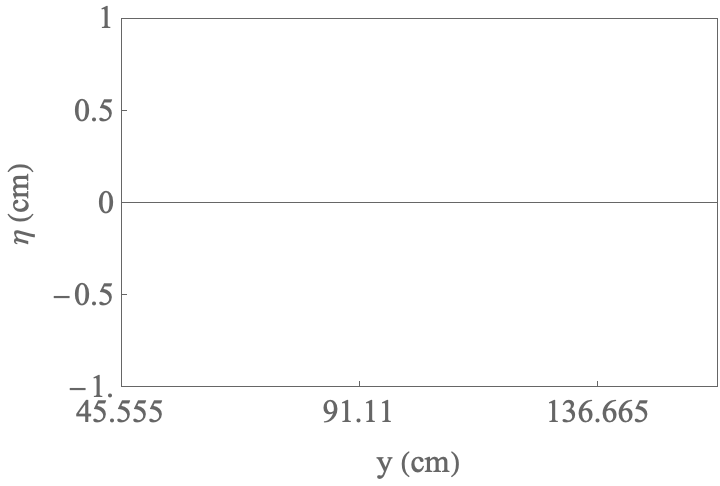}~
\includegraphics[width=1.5in]{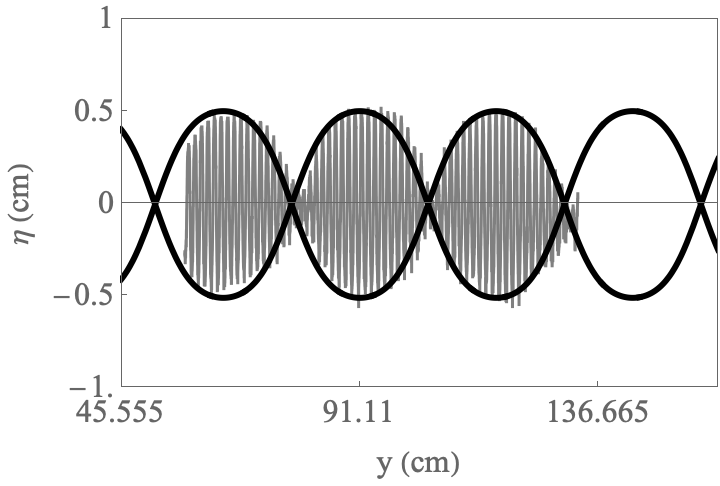}~
\includegraphics[width=1.5in]{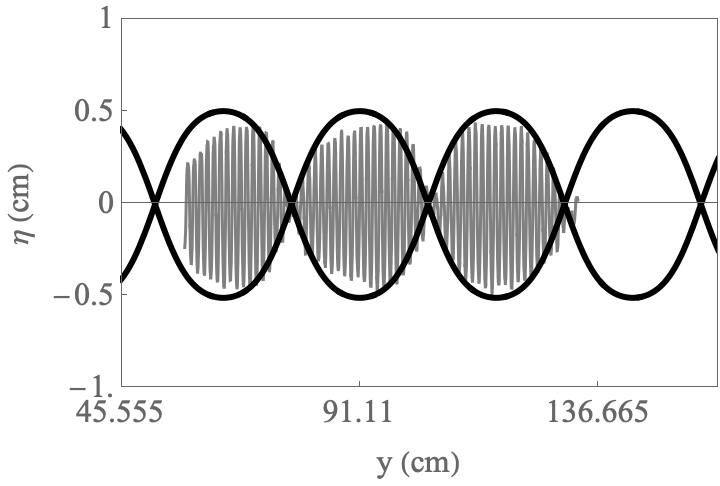}\\
\vspace{-0.2cm}
{\tiny{$n=8$}}\\

\vspace{-1.5cm}
~~~~~~~~~~~~\includegraphics[width=1.5in]{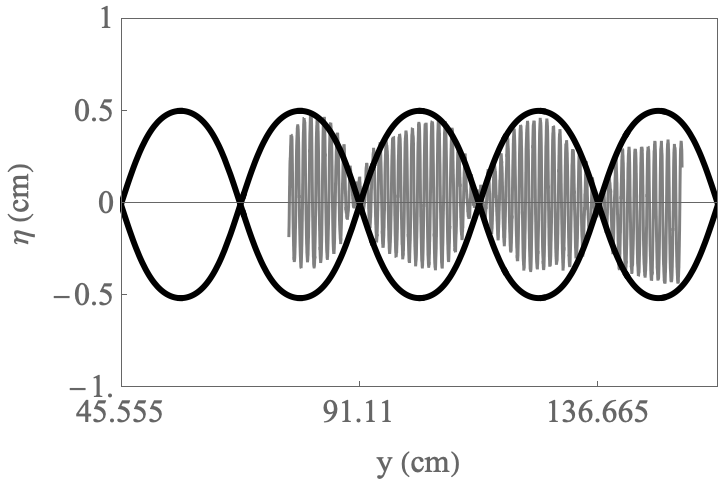}~
\includegraphics[width=1.5in]{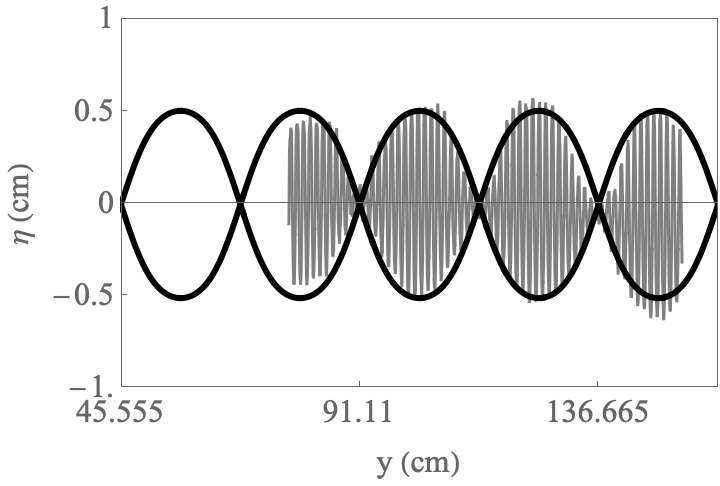}~
\includegraphics[width=1.5in]{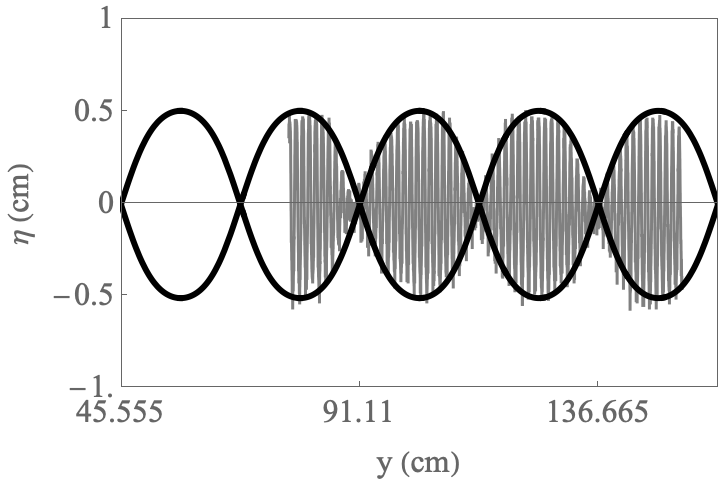}~
\includegraphics[width=1.5in]{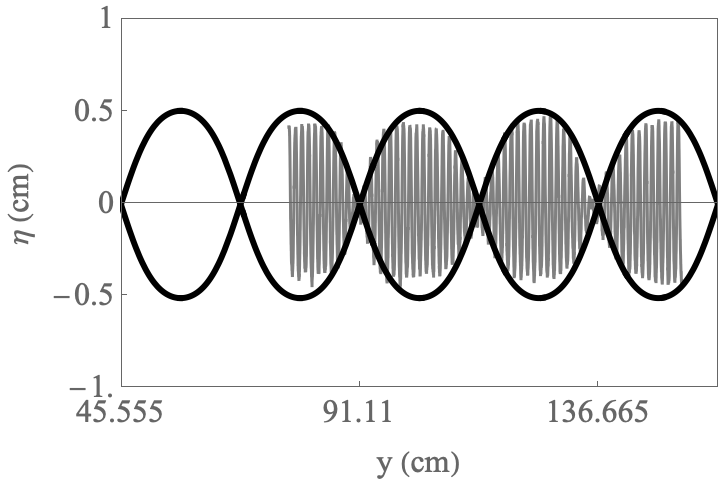}\\
\vspace{-0.2cm}
{\tiny{$n=9$}}\\

\vspace{-1.5cm}
~~~~~~~~~~~~\includegraphics[width=1.5in]{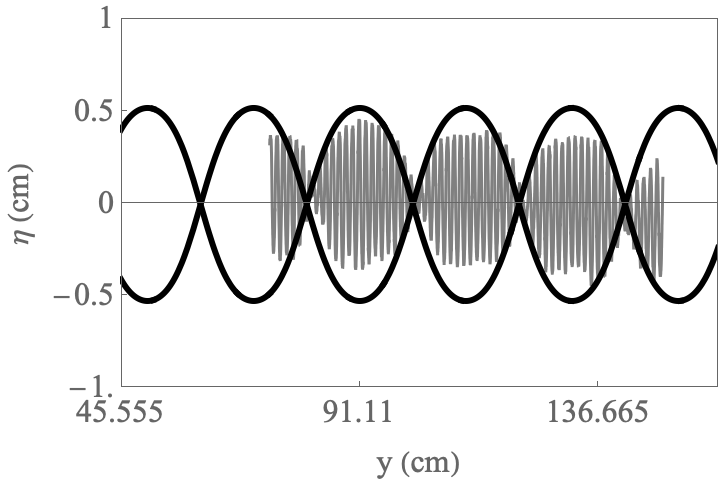}~
\includegraphics[width=1.5in]{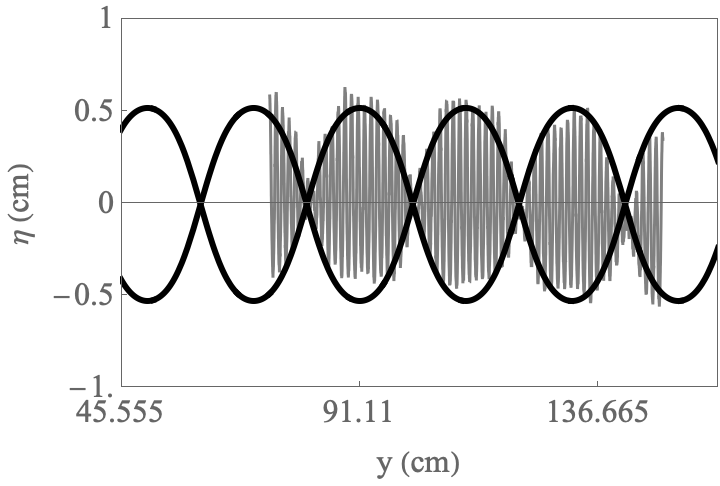}~
\includegraphics[width=1.5in]{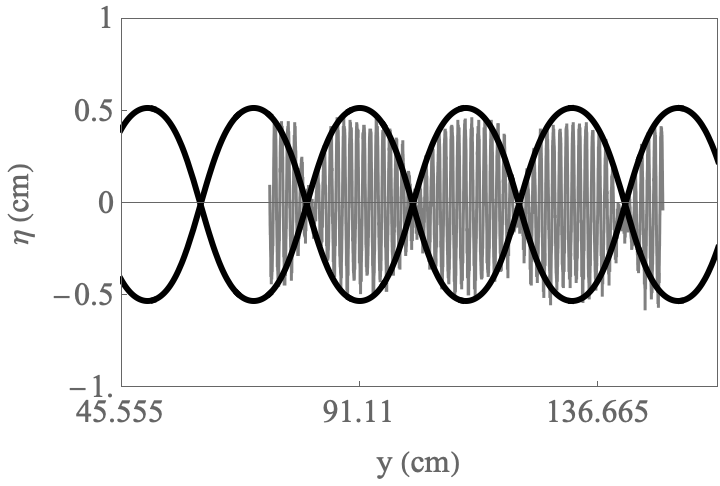}~
\includegraphics[width=1.5in]{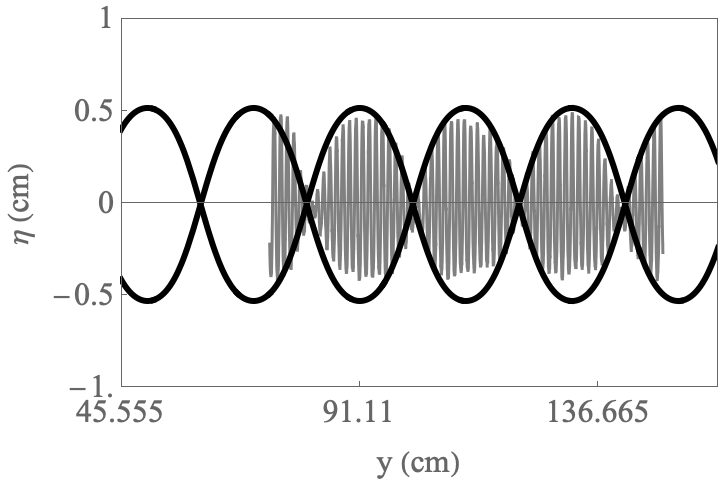}\\
\vspace{-0.2cm}
{\tiny{$n=10$}}\\

\vspace{-1.5cm}
~~~~~~~~~~~~\includegraphics[width=1.5in]{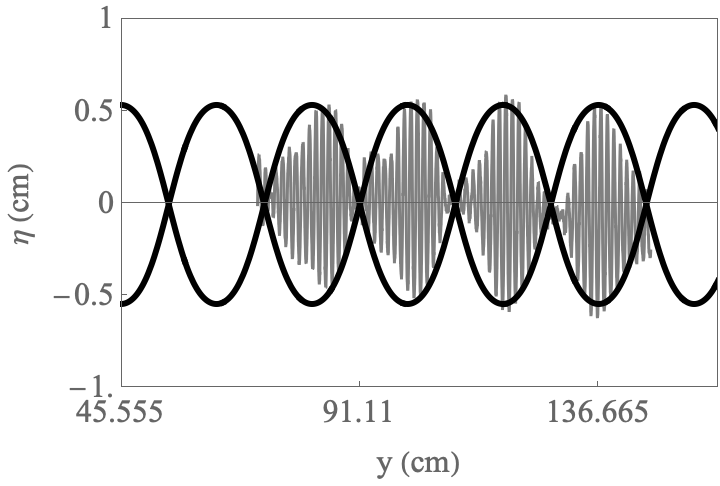}~
\includegraphics[width=1.5in]{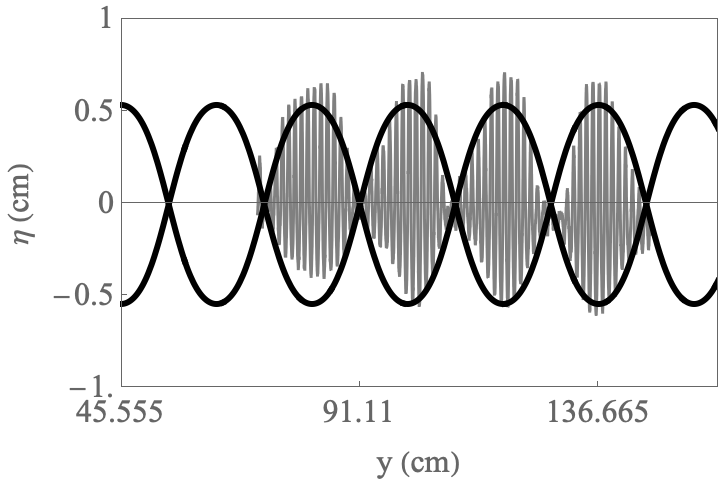}~
\includegraphics[width=1.5in]{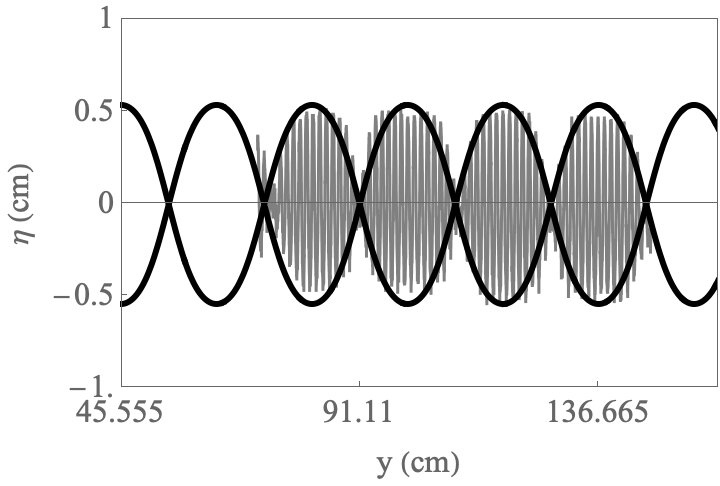}~
\includegraphics[width=1.5in]{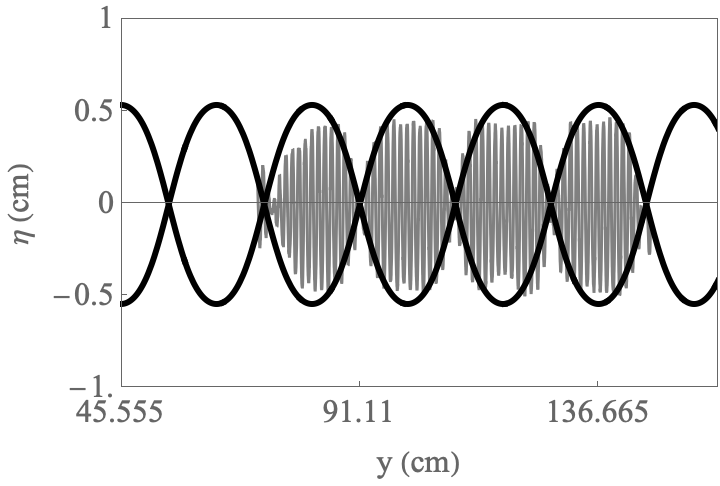}\\
\vspace{-0.2cm}
{\tiny{$n=11$}}\\

\vspace{-1.5cm}
~~~~~~~~~~~~\includegraphics[width=1.5in]{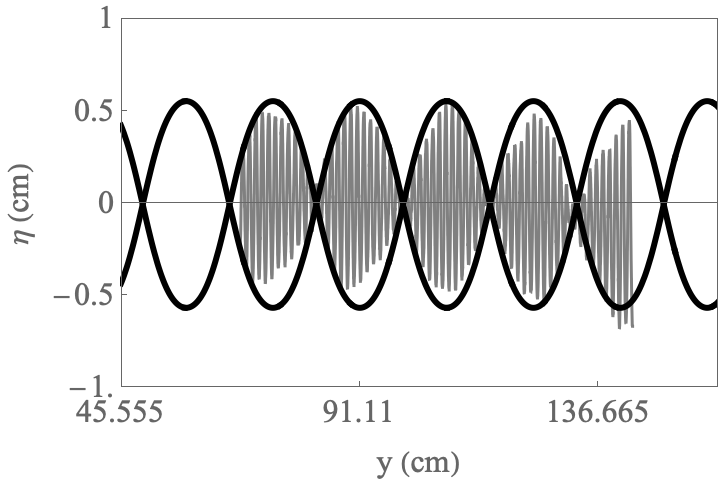}~
\includegraphics[width=1.5in]{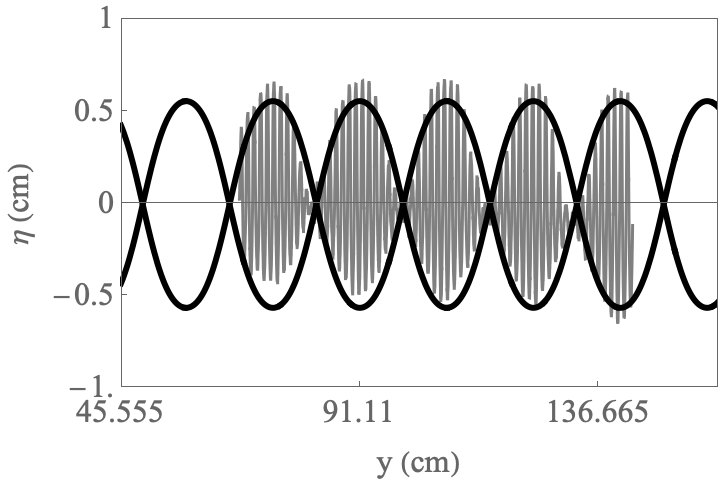}~
\includegraphics[width=1.5in]{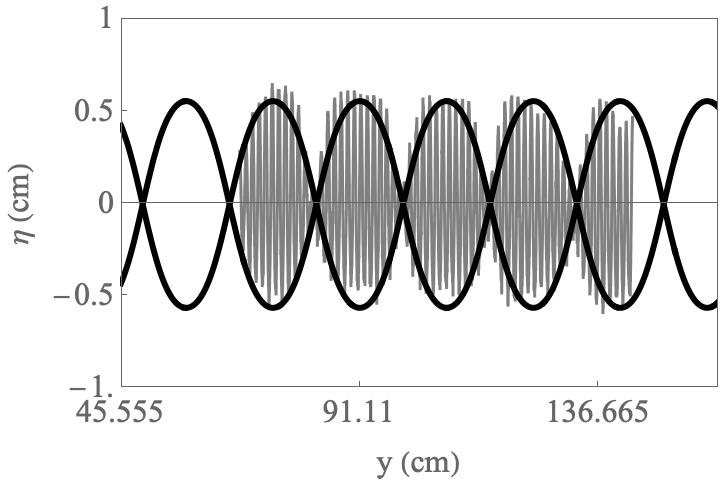}~
\includegraphics[width=1.5in]{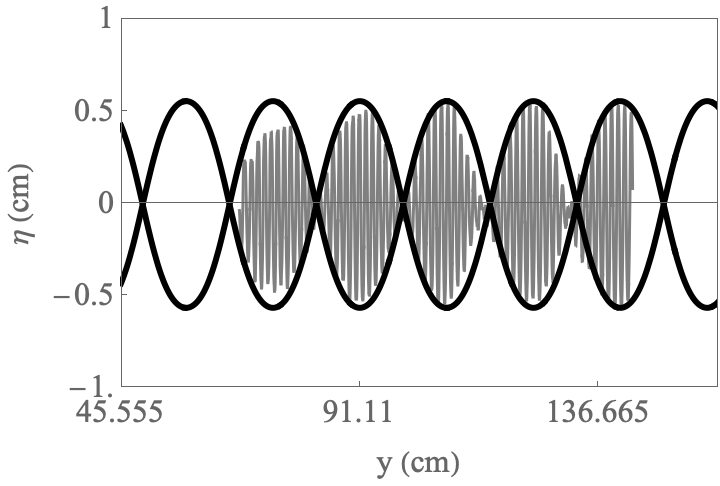}
\caption{\label{figdata} Surface displacement in the transverse direction (gray curves) in experiments with $n=4, 5, \dots, 11$ in each row as indicated for
(a) $x_1=60$ cm, (b) $x_2=90$ cm, (c) $x_3=120$ cm, (d) $x_4=150$ cm from the wave paddles; and envelopes 
(black curves) corresponding to the sn solution of the
sNLSE  given by {\ref{exactsurface}} with $x=t=0$.
Parameters for the sn solutions are given in Table~{\ref{tablesnfit}}.}
\end{figure}

\subsection{Extra amplitude variation in the $y$--direction}
\label{yenvelope}

Despite the reasonable agreement between predicted and measured amplitude variations in the $y$--direction, particularly using the sn solution from
sNLSE, the data (see Figure~{\ref{figdata}) show that there is extra, unpredicted variation of the measured amplitudes
at a fixed $x$, for increasing
values of $y$. 
Since the sn solution from sNLSE agrees best with the data, we use it as a model to investigate this extra variation.
To this end, we looked at the half-periods of the amplitude variations in the $y$--direction. For example, consider
Figure~{\ref{extraymodel}}, which shows the same data as in the $n=10$ row, 4th column of Figures~{\ref{figdata}}.
The numbers, $1,\dots,4$, label four half-periods. We computed the error for each of these half-periods using ({\ref{eqnerror}}) for all
of the experiments with more than one (full) half-period.
Table~{\ref{tableyerror}} shows the errors for the data shown in
Figure~{\ref{figdata}} for $n > 5$. 
(For $n=4, 5$ there is only one full half-period, so those experiments do not play a role here.)
The number of half-periods for experiments with $n=6, 7, 8$ was 2, for $n=9$ was 3; and for 
$n=10, 11$ was 4.
\begin{figure}
{\centerline{\includegraphics[width=2.5in]{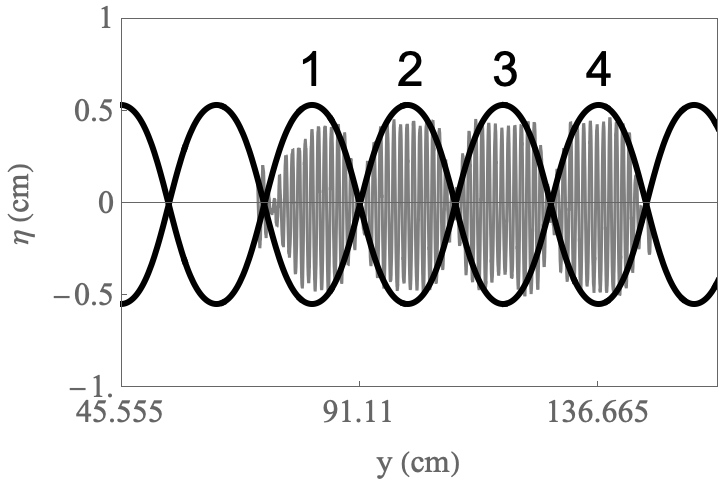}}}
\caption{\label{extraymodel} Surface displacement in the transverse direction (gray curves) in the
experiments with $n=10$ obtained at
$x_4=150$ cm. The envelopes
(black curves) corresponding to the sn solution of the
sNLSE  given by {\ref{exactsurface}} with $x=t=0$.
Parameters for the sn solutions are given in Table~{\ref{tablesnfit}}.}
\end{figure}

The comparison is made for the wave profiles obtained at the first and last
gage sites in $x$, i.e., $x=x_1$ and $x=x_4$. For $n=9, 10$, the error in Table~{\ref{tableyerror}} increases with increasing $y$ at $x=x_1$.
This increase in error would be consistent with the 
stability calculation discussed in \S{\ref{theory-snlse}} and 
Figure~{\ref{figsnstab}}. One possible explanation
for why there would be a symmetry-breaking instability like the one discussed there is that it is introduced
by the in-situ gage as it traverses the tank in a preferred direction. 
The extra amplitude variation in the $y$--direction decreases from the first to fourth gage, consistent with the previous observations and theoretical work by Segur and colleagues (\cite{bf05}, \cite{hsc10}, \cite{hs13})
 that instabilities are stabilized in the direction of propagation due to dissipation.

\begin{table}[ht]
\begin{center}
\begin{tabular}{| c ||  c | c | c  | c |  c  |c | c|  c| c |c  |c | }
\hline
$n$                  &  half-period 1     &   half-period 2     &   half-period 3     &   half-period 4        \\
 \hline
 6   ~($x_1$)      &     0.018             & 0.006           &     n/a                  &   n/a   \\
 ~  ~ ($x_4$)      &     0.014             & 0.003           &     n/a                  &   n/a   \\
 7  ~ ($x_1$)      &    n/a                  & n/a               &     n/a                  &   n/a   \\
 ~  ~ ($x_4$)      &    0.029              & 0.015           &     n/a                  &   n/a   \\
 8  ~ ($x_1$)      &     0.037             & 0.014           &     n/a                  &   n/a   \\
 ~  ~ ($x_4$)      &     0.017             & 0.011           &     n/a                  &   n/a   \\
 9  ~ ($x_1$)      &     0.026             & 0.060           &     0.089                  &   n/a   \\
 ~ ~  ($x_4$)      &     0.005             & 0.025          &     0.004                  &   n/a   \\
 10   ($x_1$)      &     0.063             & 0.086           &     0.113              &   0.103   \\
 ~  ~ ($x_4$)      &     0.053             & 0.035           &     0.052              &   0.034   \\
 11   ($x_1$)      &     0.030             & 0.018           &     0.013              &   0.025   \\
 ~  ~ ($x_4$)      &     0.004             & 0.020           &     0.014              &   0.015   \\
 \hline
\end{tabular}
\end{center}
 \caption {Error as a function of $y$ obtained at $x=x_1$ and $x=x_4$.
 The error is computed using ({\ref{eqnerror}}) in comparisons of the predictions of amplitude envelope from ({\ref{exactsurface}}) 
 (with $x=t=0$) for sequential (in $y$) half-periods of the $y$ envelopes. }
\label{tableyerror}
\end{table}

\section{Summary}
\label{summary}

Bi-periodic patterns of waves that propagate in the $x$--direction with amplitude variations in the $y$--direction are generated in the laboratory.  The variation of their amplitudes in the $y$--direction are studied within the framework of the vector (vNLSE) and scalar (sNLSE) nonlinear Schr\"odinger equations. They can be described by the uniform amplitude, Stokes-like solution of the vNLSE and the Jacobi elliptic
sine (sn) function solution of the sNLSE.  Our main results are the following.
    \begin{enumerate}
    \item The errors between predictions and measurements increase with increasing $\epsilon_y$, the measure of two-dimensionality of the wave patterns. This result might be anticipated for the predictions from the sNLSE model, which assumes that  $\epsilon_y \ll 1$. But it was also true for predictions from the vNLSE model, which makes no constraint on $\epsilon_y$.
    \item The wave patterns were generated using the Stokes-type solution of the vNLSE equation. Therefore, one might anticipate that that model would provide the best description of the measured wave patterns. It did not. The errors between the sn solution of the sNLSE were  less than those of the Stokes-type solution of the vNLSE. 
    \item The amplitude variation in the $y$--direction of the measured wave patterns is not sinusoidal, especially for $\epsilon_y \lessapprox 0.25$. The $y$--envelope is flattened. To account for this flattening, the Stokes-type solution of the vNLSE requires the inclusion of a third-harmonic (in the measure of weak nonlinearity) term. The sn solution of the sNLSE accounts for the flattening
    directly through the elliptic modulus. Therefore to describe the wave patterns, the sNLSE model requires a single mode with no higher-order terms. 
    \end{enumerate}
The measurements do not show modulations or evidence of instabilities in the direction of propagation, consistent with the work of Segur and colleagues, who showed that dissipation stabilizes the modulational instability. There is some extra, unpredicted variability in the amplitude variation in the $y$--direction that increases in $y$ at the first gage site.  
This increase may be consistent with a qualitative stability calculation that allows symmetry breaking in that direction.

\section{Acknowledgements}

This material is based upon work supported by the National Science Foundation under Grant Nos. DMS-1716159 (DMH) and DMS-1716120 (JDC).

\appendix
\section{Formula for $\zeta$ including surface tension}
\label{append}

The formula for $\zeta$ including surface tension, see equation (\ref{zeta-vnlse}), is 

\begin{align*}
  \zeta=&\Bigg{(}-8 g^4 (5 k^6-k^4 l^2-11 k^2 l^4-4 l^6-4 k^5 \kappa +4 k^3 l^2 \kappa +8 k l^4 \kappa )\\
  &+g^3 \Big{(}-213 k^8-251 k^6
  l^2+461 k^4 l^4+655 k^2 l^6+188 l^8+76 k (3 k^2-5 l^2) \kappa ^5\Big{)} \sigma\\
  &-4 g^2 \Big{(}99 k^{10}+311 k^8 l^2-69 k^6 l^4-577 k^4 l^6-398 k^2 l^8-102 l^{10}-99 k^9 \kappa -280 k^7 l^2 \kappa \\
  & \hspace{2cm}         +224 k^5 l^4 \kappa +492 k^3 l^6 \kappa +183 k l^8 \kappa \Big{)} \sigma ^2\\
  &-4 g \Big{(}49 k^{12}+494 k^{10} l^2+319 k^8 l^4-644 k^6 l^6-833 k^4 l^8-410 k^2 l^{10}-95 l^{12} \\
  & \hspace{2cm}   +2 k (-25 k^{10}-231 k^8 l^2-63 k^6 l^4+327
  k^4 l^6+316 k^2 l^8+68 l^{10}) \kappa\Big{)} \sigma ^3\\
  &+16 l^2 \kappa ^2 \Big{(}-49 k^{10}-44 k^8 l^2+74 k^6 l^4+60 k^4 l^6+31 k^2 l^8+8 l^{10}+2 k (25 k^8+6 k^6 l^2-27 k^4 l^4-28 k^2 l^6-4
  l^8) \kappa\Big{)} \sigma ^4\Bigg{)} \\
  &/\Bigg{(}8 \kappa (g+4 l^2 \sigma )(g (k-2 \kappa )+4 k^3 \sigma -2 \kappa ^3 \sigma)^2\omega\Bigg{)}
\end{align*}

\bigskip


\end{document}